\documentclass[preprint2,tighten]{aastex}
\usepackage[]{times, graphicx}
\newcommand{\pref}{\protect\ref}

\shorttitle{\indent \def Spectroscopic observations of CMEs, dimmings, and jets} \shortauthors{Tian et al.}

\begin{document}

\title{What can we learn about solar coronal mass ejections, coronal dimmings, and Extreme-Ultraviolet jets through spectroscopic observations?}

\author{Hui Tian\altaffilmark{1}, Scott W. McIntosh\altaffilmark{1}, Lidong Xia\altaffilmark{2}, Jiansen He\altaffilmark{3}, Xin Wang\altaffilmark{1,3}}
\altaffiltext{1}{High Altitude Observatory, National Center for Atmospheric Research, P.O. Box 3000, Boulder, CO 80307; htian@ucar.edu}
\altaffiltext{2}{Shandong Provincial Key Laboratory of Optical Astronomy and Solar-Terrestrial Environment, School of Space Science and Physics, Shandong University at Weihai, Weihai 264209, China}
\altaffiltext{3}{School of Earth and Space Sciences, Peking University, China}

\begin{abstract}
Solar eruptions, particularly coronal mass ejections (CMEs) and extreme-ultraviolet (EUV) jets, have rarely been investigated with spectroscopic observations. We analyze several data sets obtained by the EUV Imaging Spectrometer onboard {\it Hinode} and find various types of flows during CMEs and jet eruptions. CME-induced dimming regions are found to be characterized by significant blueshift and enhanced line width by using a single Gaussian fit. While a red-blue (RB) asymmetry analysis and a RB-guided double Gaussian fit of the coronal line profiles indicate that these are likely caused by the superposition of a strong background emission component and a relatively weak ($\sim$10\%) high-speed ($\sim$100~km~s$^{-1}$) upflow component. This finding suggests that the outflow velocity in the dimming region is probably of the order of 100~km~s$^{-1}$, not $\sim$20~km~s$^{-1}$ as reported previously. Density and temperature diagnostics of the dimming region suggest that dimming is primarily an effect of density decrease rather than temperature change. The mass losses in dimming regions as estimated from different methods are roughly consistent with each other and they are 20\%-60\% of the masses of the associated CMEs. With the guide of RB asymmetry analysis, we also find several temperature-dependent outflows (speed increases with temperature) immediately outside the (deepest) dimming region. These outflows may be evaporation flows which are caused by the enhanced thermal conduction or nonthermal electron beams along reconnecting field lines, or induced by the interaction between the opened field lines in the dimming region and the closed loops in the surrounding plage region. In an erupted CME loop and an EUV jet, profiles of emission lines formed at coronal and transition region temperatures are found to exhibit two well-separated components, an almost stationary component accounting for the background emission and a highly blueshifted ($\sim$200~km~s$^{-1}$) component representing emission from the erupting material. The two components can easily be decomposed through a double Gaussian fit and we can diagnose the electron density, temperature and mass of the ejecta. Combining the speed of the blueshifted component and the projected speed of the erupting material derived from simultaneous imaging observations, we can calculate the real speed of the ejecta.
\end{abstract}


\section{Introduction}
Coronal mass ejections (CMEs) are large-scale solar eruptions and earth-directed CMEs are often sources of strong geomagnetic storms \citep[e.g.,][]{Gosling1991,Wang2002,Wang2006,Zhang2005,Feng2009,Liu2010}. Recent statistical studies of \cite{Reinard2008} and \cite{Bewsher2008} have shown that more than 50\% of the frontside CMEs are associated with coronal dimmings (or transient coronal holes), which are characterized by abruptly reduced emission in extreme-ultraviolet (EUV) and soft X-rays \citep[e.g.,][]{Rust1976,Gopalswamy1998,Thompson1998,Zarro1999,Zhou2003,DeToma2005,McIntosh2007,Miklenic2011}. Dimmings may mark locations of the footpoints of ejected flux ropes \citep[e.g.,][]{Webb2000} or formed by reconnection between the erupting field and the surrounding magnetic structures \citep[e.g.,][]{Attrill2007,Mandrini2007}. There are basically two types of dimmings: small-scale dimmings associated with the two ends of a pre-CME sigmoid structure \citep[e.g.,][]{Sterling1997,Hudson1998,Zarro1999,Webb2000,Jiang2003,Cheng2010} and global-scale dimmings which are often immediately proceeded by global "EUV waves" \citep[e.g.,][]{Thompson2000,Attrill2007}. 

Jet-like phenomena are small-scale solar eruptions and they are often observed in X-ray, EUV, and white light. Most jets are associated with small flares \citep{Madjarska2007}. EUV jets are characterized by nearly collimated high-speed motions of plasma at coronal and transition region (TR) temperatures \citep[e.g.,][]{Alexander1999,Lin2006,Liu2011,Shen2011,Shen2012,Srivastava2011}. Studies have shown that EUV jets and X-ray jets are closely associated with each other \citep{Kim2007,Chifor2008b,He2010b,YangL2011}. Recent observations by the Atmospheric Imaging Assembly \citep[AIA,][]{Lemen2011} onboard the Solar Dynamics Observatory ({\it SDO}) have revealed that fine-scale EUV jets (high-speed outflows) are ubiquitous on the Sun \citep{DePontieu2011,Tian2011b,YangS2011}.

Kinematics associated with CMEs and EUV jets are usually studied through coronagraph and broadband observations. The high cadence and large field of view (FOV) of these imaging observations have greatly enhanced our understanding of these solar eruptions. However, imaging observations only allow us to study the plane of sky (POS) component of the kinematics, which is usually a good approximation of the full kinematics only for limb events. For earth-directed eruptions, especially halo-CMEs which are the cause of most strong geomagnetic storms, imaging instruments placed close to the Sun-Earth line often fail to observe their initial or complete evolution. Spectroscopic observations, on the other hand, can provide information on the plasma motions in the line of sight (LOS) direction and thus are critical for us to understand the kinematics of earth-directed eruptions. For both disk and limb eruptions, their three-dimensional (3-D) evolution can in principle be revealed through simultaneous imaging and spectroscopic observations. In addition, spectra of different emission lines can be used to diagnose plasma properties such as electron density and temperature. Spectroscopic data can also be used to estimate the mass of the erupted material and mass loss in the dimming region. 

So far there are only a few spectroscopic investigations of CMEs, dimmings, and EUV jets in the literature. Using observations by the Coronal Diagnostic Spectrometer \citep[CDS,][]{Harrison1995} onboard the Solar and Heliospheric Observatory (SOHO), \cite{Harra2001} reported significant blue shift of emission lines formed at coronal and TR temperatures in dimming regions. This result has been confirmed by recent high-resolution observations of the EUV Imaging Spectrometer \citep[EIS,][]{Culhane2007} onboard {\it Hinode} \citep{Harra2007,Jin2009,Attrill2010,Chen2010a,Harra2011a}. Blue shift was also found in footpoint regions of small-scale erupted loops \citep{He2010b}. However, a preliminary study of \cite{McIntosh2010} suggests that some line profiles in the dimming regions are asymmetric, with a weak enhancement in the blue wings. EIS observations have also revealed an obvious increase of the line broadening in dimming regions, which was interpreted as a growth of Alfv\'en wave amplitude or inhomogeneities of flow velocities along the LOS \citep{McIntosh2009c,Chen2010a,Dolla2011}. The presence of asymmetric line profiles suggests that there are probably two emission components and that a single Gaussian fit may not reveal the real physics in dimming regions \citep{McIntosh2010,Dolla2011}.

The outflow speed derived from a single Gaussian fit is roughly in the range of 10-40~km~s$^{-1}$ and usually it does not change significantly for coronal emission lines formed at different temperatures \citep{Harra2007,Jin2009,Attrill2010,Chen2010a}. However, \cite{Imada2007} reported a temperature-dependent outflow in the dimming region following a CME. The speed of the flow increases from $\sim$10~km~s$^{-1}$ at log ({\it T}/K)=4.9 to $\sim$150~km~s$^{-1}$ at log ({\it T}/K)=6.3. One-dimensional modeling effort has been taken to reconstruct this temperature-dependent outflow \citep{Imada2011}.

Line splitting is usually associated with a very high-speed ($\sim$200~km~s$^{-1}$ or larger) plasma motion. Using CDS and EIS observations, \cite{Harra2003}, \cite{Asai2008} and \cite{Li2012} found signatures of line splitting indicative of plasma ejection at a speed of $\sim$250~km~s$^{-1}$ during CMEs or filament eruptions. Spectra obtained by the Solar Ultraviolet Measurements of Emitted Radiation Spectrograph \citep[SUMER,][]{Wilhelm1995,Lemaire1997} onboard SOHO have revealed signatures of line splitting associated with the expanding X-ray plasma in a flare/CME event \citep{Innes2001}. Line splitting or obviously blueshifted components have also been found in spectra of EUV jets in coronal holes and ARs \citep{Wilhelm2002,Madjarska2007,Kamio2007,Kamio2009,Chifor2008a}. 

There have been a few investigations of the plasma properties of dimmings and EUV jets. \cite{Harrison2000} and \cite{Harrison2003} used the Si~{\sc{x}}~347.40\AA{}~\&~356.04\AA{} line pairs to diagnose the electron density and found that it decreased as dimming occurred. Using some assumptions of the emitting volume and the distribution of the amount of material at different temperatures, they also made an effort to estimate the mass loss in the dimming region and found that it is of the same order as the mass of the associated CMEs. Taking values of the formation heights of different emission lines and the densities from static solar atmosphere models, \cite{Jin2009} also developed a method to estimate the mass losses in dimming regions associated with two events during 2006 Dec 13-15. Using the Fe~{\sc{xii}}~186.88\AA{}~\&~195.12\AA{} line pair, \cite{Chifor2008a} measured electron densities higher than log ({\it}$N_{e}$/cm$^{-3}$)=11 for an EUV jet. However, they only simply summed up the spectral intensities in the wavelength windows of the lines and could not separate the blueshifted component from the background emission component.

Besides dimmings and ejecta associated with CMEs and EUV jets, other solar eruption related phenomena such as flare induced chromospheric evaporation \citep{Teriaca2003,Milligan2006a,Milligan2006b,Milligan2009,Milligan2011,Chen2010b,Watanabe2010,Li2011,Graham2011,LiuR2010}, flare-related magnetic reconnection \citep{Wang2007,Hara2011}, filament oscillations \citep{Chen2008,Bocchialini2011} and coronal waves \citep{Harra2011b,Chen2011,Veronig2011} have also been investigated through EUV spectroscopic observations. As the approaching of the new solar maximum, there is no doubt that more spectroscopic observations will be employed to study solar eruptions since the high-resolution EIS instrument is still in good condition and the Interface Region Imaging Spectrograph (IRIS) is expected to be launched in 2012. 

In this paper we analyze several data sets obtained by EIS during CME eruptions and EUV jets. The shapes of the EIS spectral line profiles suggest that the emission often consists of at least two components so that previous results based on a single Gaussian fit may need to be reconsidered. We apply the recently improved techniques of red-blue (RB) asymmetry analysis and RB-guided double Gaussian fit \citep{Tian2011c}, which we used previously to study properties of the high-speed outflows in non-eruptive active regions (ARs), to the spectra acquired during solar eruptions. We find various types of flows and discuss possible mechanisms to produce these flows. We also diagnose the density, temperature and mass loss (mass) of the dimming region as well as the ejected material. Our analyses demonstrate that EUV spectroscopic observations can provide a lot of valuable information on solar eruptions.

\section{Observations, Single Gaussian fit, and RB asymmetry analysis}

\begin{table*}[]
\caption[]{EIS Observations Used in This Study} \label{table1}
\begin{center}
\begin{tabular}{p{0.6cm} | p{3.8cm} | p{1.5cm} | p{0.6cm} | p{3.5cm} | p{2.7cm}}
\hline Obs. ID & Scanning Period & Exposure time (s) & Slit &  Flare Class \& Peak Time & Comment \\
\hline 
1 & 2006 Dec 14 15:11-16:01 & 10 & $1^{\prime\prime}$ & X1.5, Dec 14 22:15 & CME \& Dimming \\
   & 2006 Dec 14 19:20-21:35 & 30 & &  &  \\
   & 2006 Dec 15 01:15-03:30 & 30 & &  &  \\
   & 2006 Dec 15 04:10-06:25 & 30 & &  &  \\
   & 2006 Dec 15 10:29-11:19 & 10 & &  &  \\
\hline
2 & 2007 May 19 09:42-10:30 & 10 & $1^{\prime\prime}$ & B9.5, May 19 13:02 & CME \& Dimming \\
   & 2007 May 19 11:41-15:23 & 40 & & &  \\
\hline
3 & 2006 Dec 12 19:07-23:46 & 30 & $1^{\prime\prime}$  & X3.4, Dec 13 02:40 & CME \& Dimming \\
   & 2006 Dec 13 01:12-05:41 & 30 & & &  \\
\hline
4 & 2011 Jun 21 02:11-05:18 & 9 & $2^{\prime\prime}$  & C7.7, Jun 21 03:25 & CME \& Dimming \\
\hline
5 & 2011 Feb 14 19:13-20:06 & 8 & $2^{\prime\prime}$  & C6.6, Feb 14 19:30 & CME  \\
\hline
6 & 2007 Jun 5 01:51-02:04 & 5 & $2^{\prime\prime}$  & C1.2, Jun 5 04:23 & EUV jet \\
   & 2007 Jun 5 04:16-04:29 & 5 & & &  \\
\hline
\end{tabular}
\end{center}
\end{table*}

Table~\pref{table1} lists some of the observation details of the six events we analyzed. The class and peak time of the associated flare are also listed for each event. There were many fast repetitive rasters (with a scanning cadence of $\sim$6 minutes) for Events 4\&5 and we only present results for several of them in this paper.

\begin{figure*}
\centering {\includegraphics[width=0.98\textwidth]{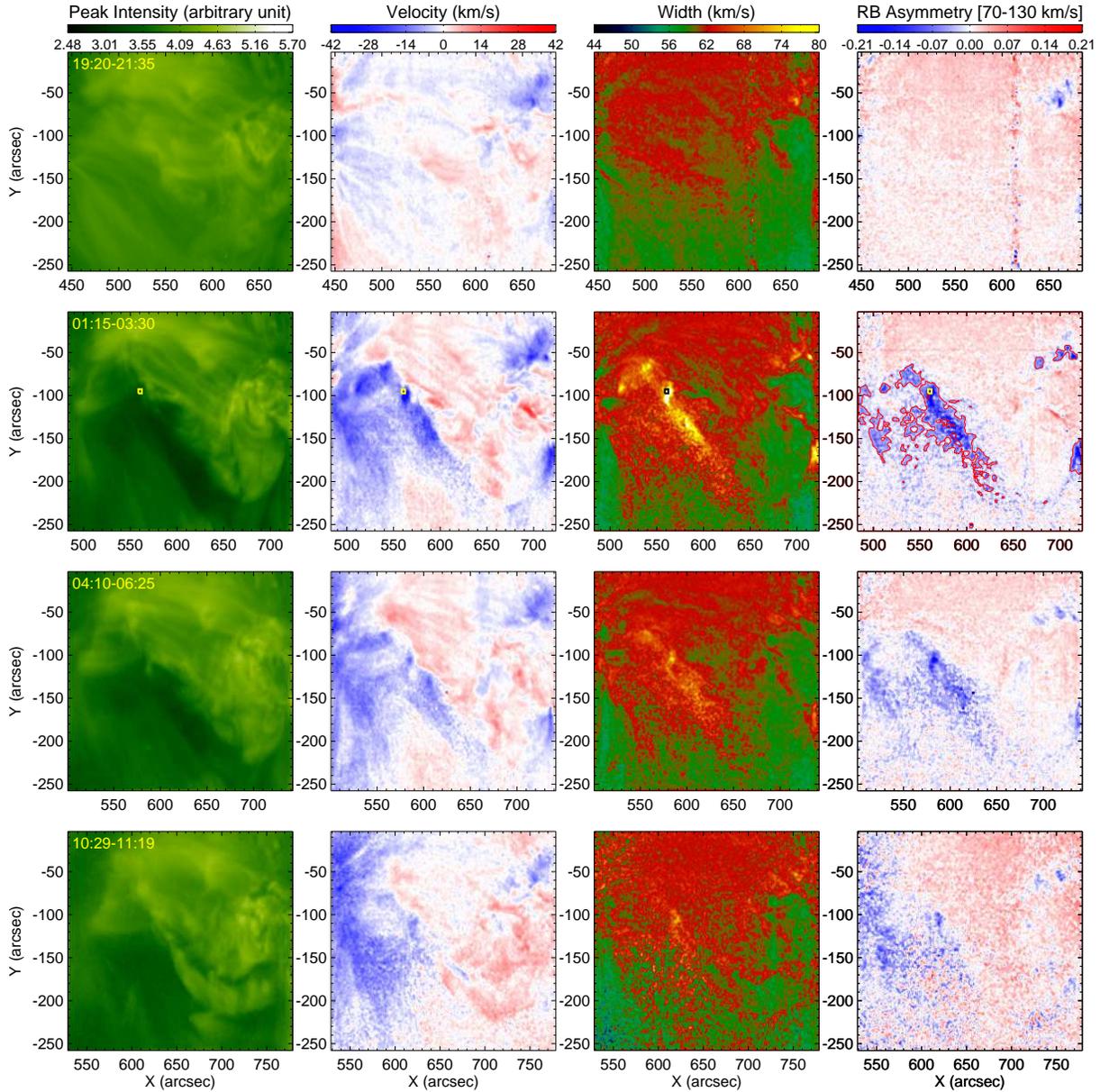}} \caption{Spatial distributions of the peak intensity, Doppler velocity and exponential width derived from the single Gaussian fit, and the average RB$_{P}$ asymmetry in the velocity interval of 70-130~km~s$^{-1}$ for Fe~{\sc{xiii}}~202.04\AA{} in the 2006 Dec 14-15 observations. The beginning and ending time (hour:minute) of each scan is indicated in the intensity image. The pre-eruption conditions are shown in the first row. The square in each panel of the second row marks locations where profiles are averaged and presented in Figure~\pref{fig.7}. The red contours shown in the map of RB$_{P}$ asymmetry for the 01:15-03:30 scan outline locations where the RB$_{P}$ asymmetry (70-130~km~s$^{-1}$) is smaller than -0.03 and the signal to noise ratio of the profile is larger than 8.} \label{fig.1}
\end{figure*}

\begin{figure*}
\centering {\includegraphics[width=0.98\textwidth]{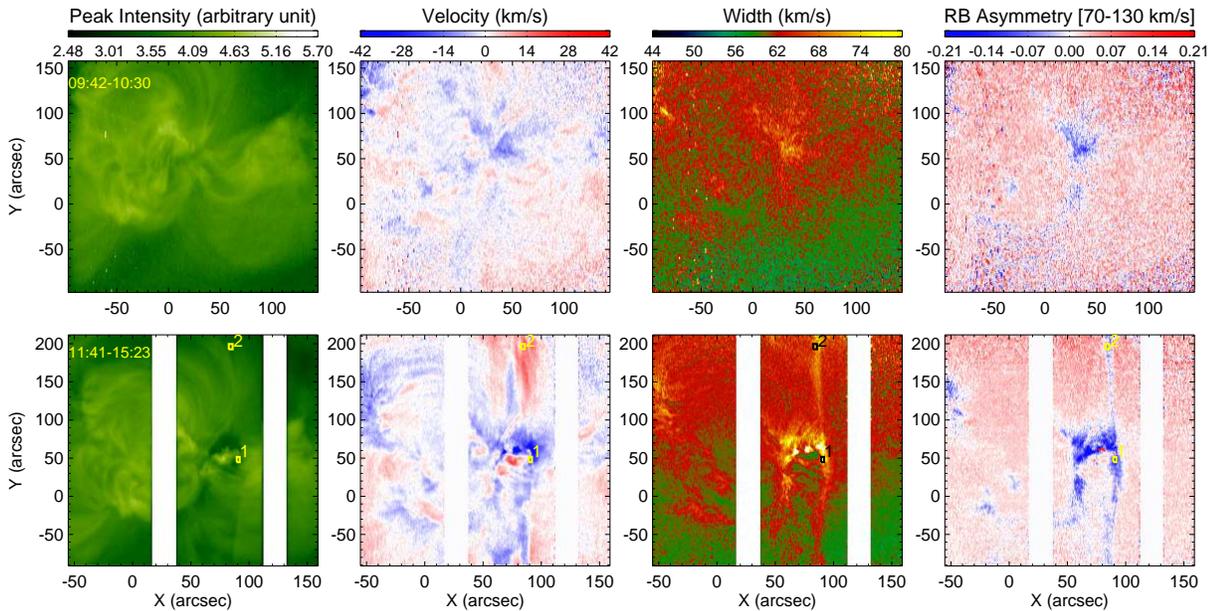}} \caption{Same as Figure~\pref{fig.1} but for the 2007 May 19 observations. The rectangular regions 1 \& 2 mark the locations where the Fe~{\sc{xiii}}~202.04\AA{} profiles are averaged and presented in Figure~\pref{fig.8}. } \label{fig.2}
\end{figure*}

\begin{figure*}
\centering {\includegraphics[width=0.98\textwidth]{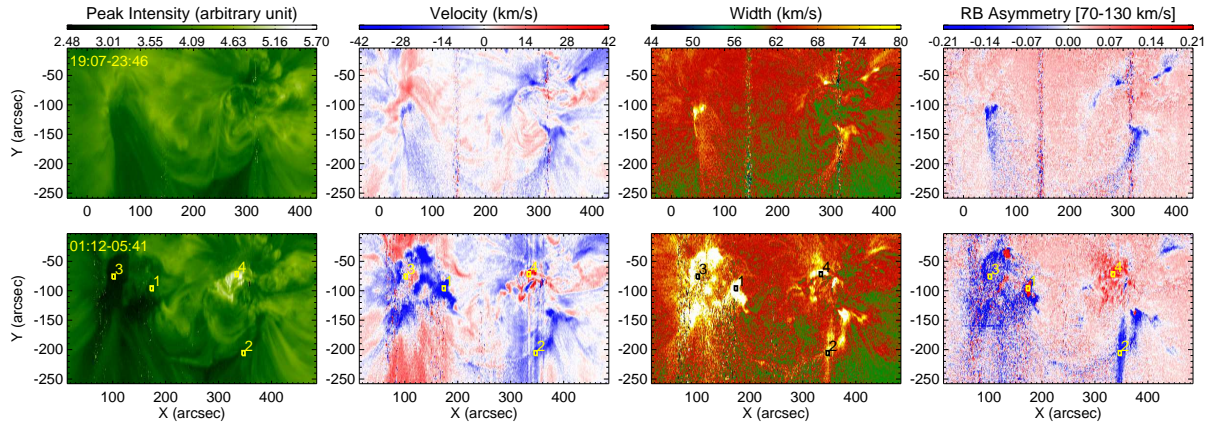}} \caption{Same as Figure~\pref{fig.1} but for the 2006 Dec 12-13 observations. The rectangular regions 1-4 mark the locations where line profiles are averaged and presented in Figures~\pref{fig.10}\&\pref{fig.14}. } \label{fig.3}
\end{figure*}

\begin{figure*}
\centering {\includegraphics[width=0.98\textwidth]{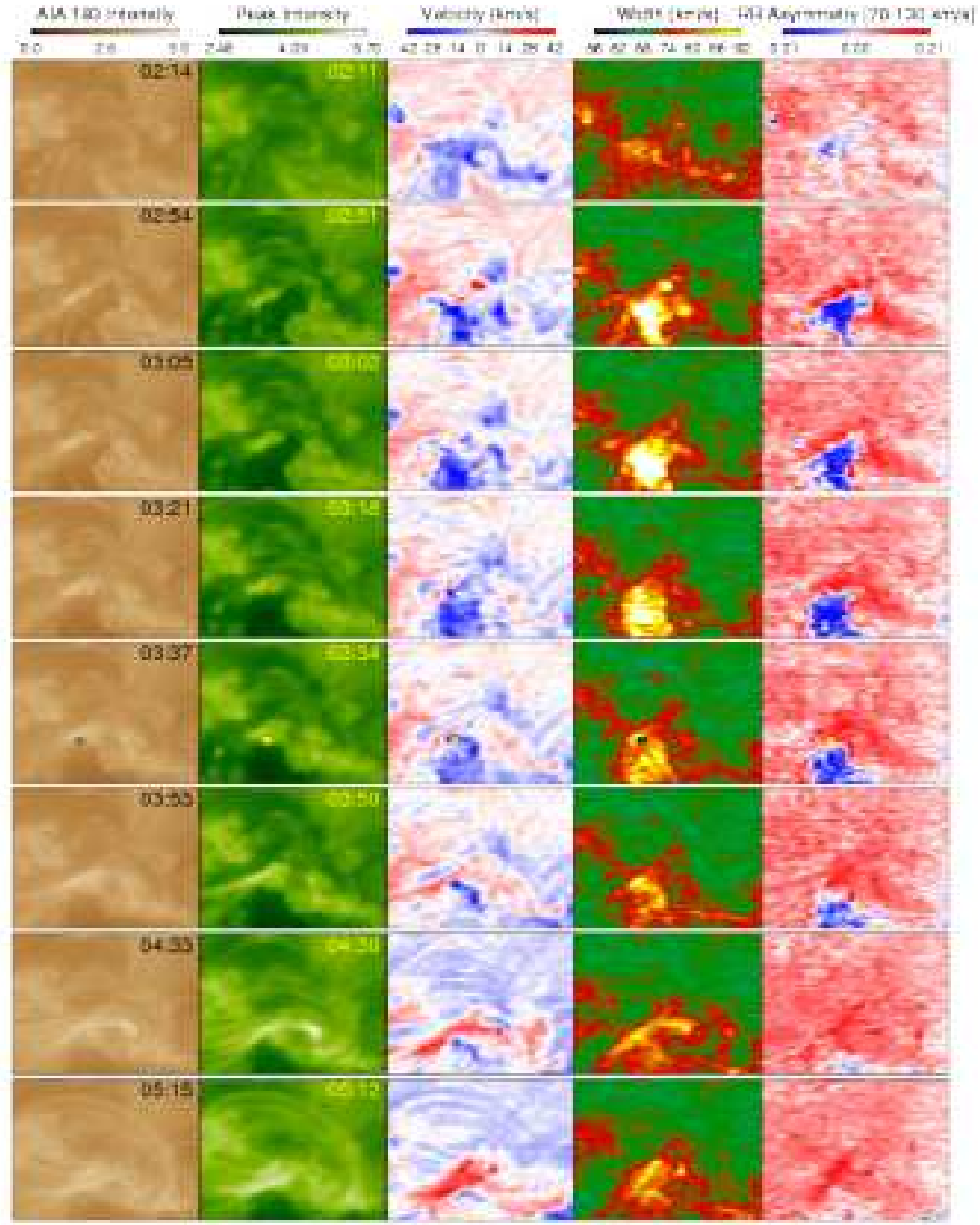}} \caption{Evolution of AIA 193\AA{} intensity and EIS Fe~{\sc{xii}}~195.12\AA{} line parameters (peak intensity, velocity and width derived from the single Gaussian fit, and the average RB$_{P}$ asymmetry in the velocity interval of 70-130~km~s$^{-1}$) in the 2011 Jun 21 observations. The time of the AIA observation and the beginning time of each EIS scan are indicated in the corresponding intensity images. The rectangular region marks the locations where line profiles are averaged and presented in Figure~\pref{fig.11}. The size of the FOV is about $175^{\prime\prime}\times152^{\prime\prime}$. A movie (m4.mpeg) showing the evolution of AIA 171\AA{} is available online. } \label{fig.4}
\end{figure*}

\begin{figure*}
\centering {\includegraphics[width=0.98\textwidth]{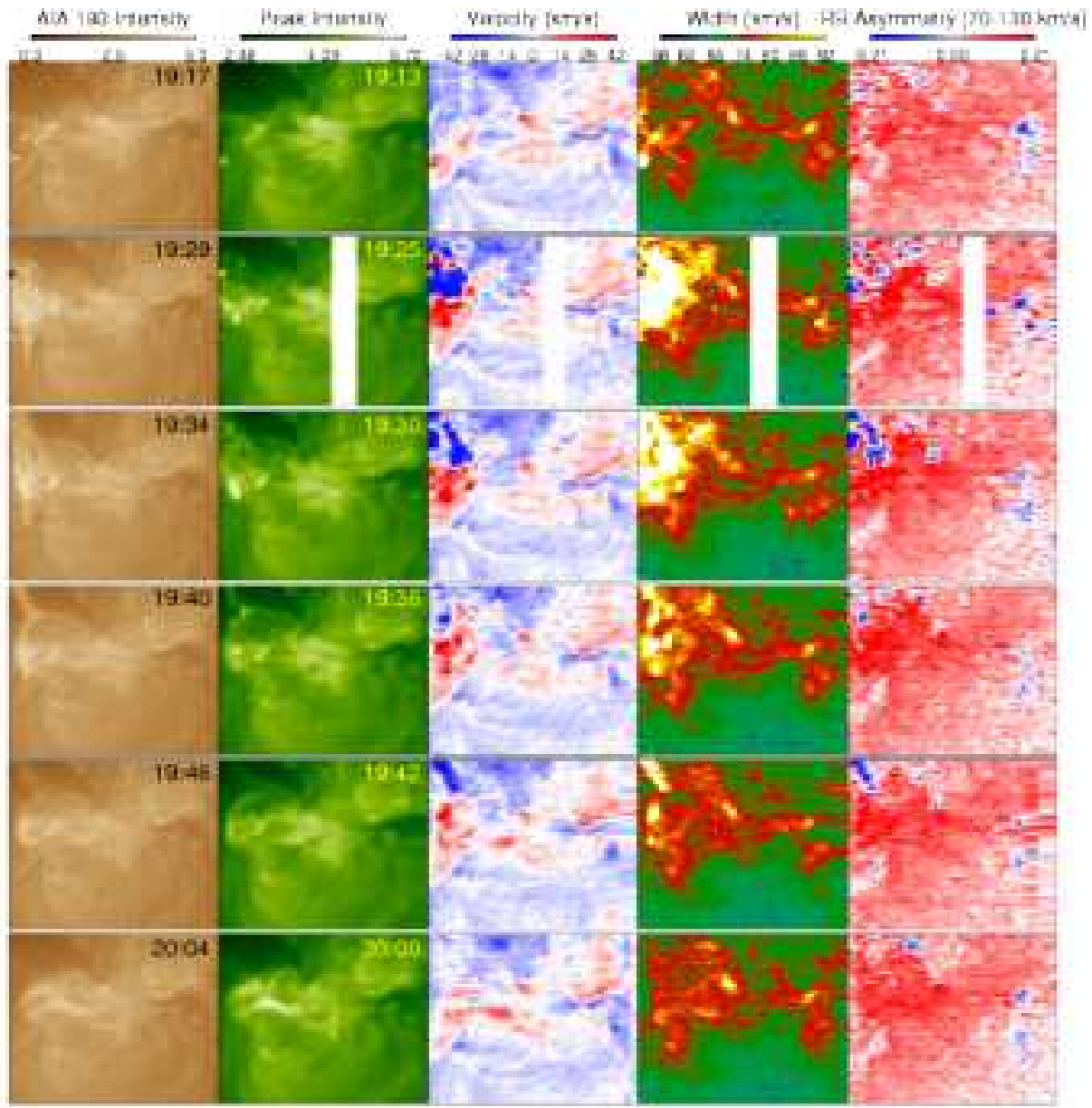}} \caption{Same as Figure~\pref{fig.4} but for the 2011 Feb 14 observations. The asterisk marks the center of five pixels along the slit where line profiles are averaged and presented in Figure~\pref{fig.12}. The size of the FOV is about $175^{\prime\prime}\times160^{\prime\prime}$. For illustration some bad data from single exposures are replaced by the data of adjacent exposures. A movie (m5.mpeg) showing the evolution of AIA 193\AA{} is available online. } \label{fig.5}
\end{figure*}

\begin{figure*}
\centering {\includegraphics[width=0.98\textwidth]{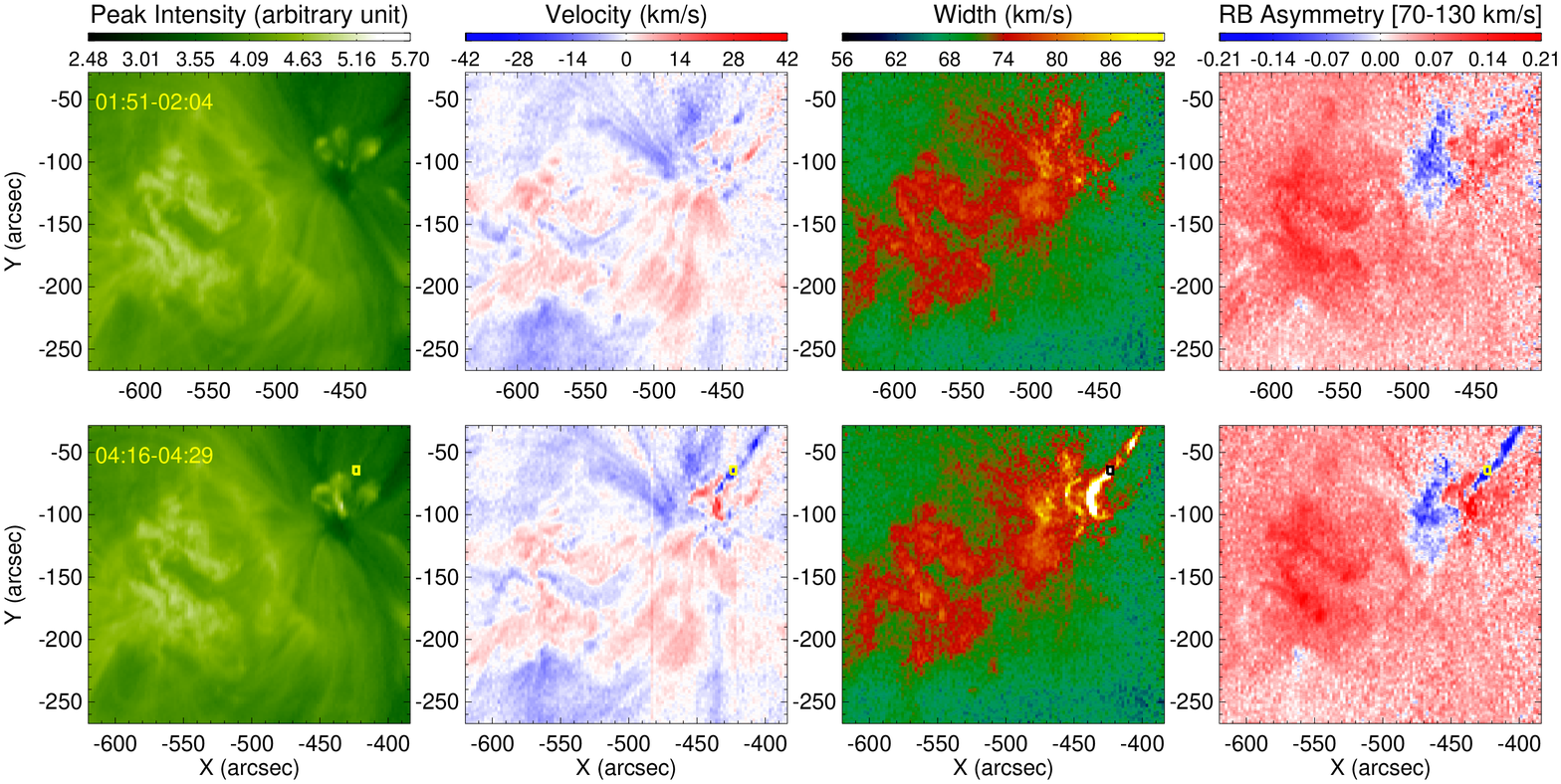}} \caption{Same as Figure~\pref{fig.1} but for Fe~{\sc{xii}}~195.12\AA{} in the 2007 Jun 5 observations. The rectangular region marks the locations where line profiles are averaged and presented in Figure~\pref{fig.13}. } \label{fig.6}
\end{figure*}

The SSW routine {\it eis\_prep.pro} was applied to correct and calibrate the EIS data. This includes CCD pedestal and dark current subtraction,
cosmic ray removal, warm and hot pixels identification, absolute calibration, error estimation, and so on. The effects of slit tilt and orbital
variation (thermal drift) were estimated by using the SSW routine {\it eis\_wave\_corr.pro} and removed from the data. After that, a running
average over 3 pixels along the slit was applied to the spectra to improve the signal to noise ratio. Note that \cite{Tian2011c} used the median values of the measurement errors when averaging profiles over several pixels. In this paper we regard these line profiles as independent measurements of the profile at a single pixel and use the uncertainty propagation theory to calculate the measurement errors for the averaged profile. This usually leads to smaller values of the errors, which is reasonable since the spatial average improves the signal to noise ratio. 

As a common practice, a single Gaussian fit was applied to each spectrum. The line peak intensity, Doppler shift and line width can thus be derived. We assume zero shift of the profile averaged over each observation region. We have to mention that the line width can be expressed in different formats and different names are assigned to different formats \citep[e.g.,][]{Chae1998,Peter2010}. A Gaussian line profile can be expressed as:

\begin{equation}
\emph{$I(v)=I_p~exp(-\frac{1}{2}\frac{(v-v_0)^2}{\sigma^2})$}\label{equation1},
\end{equation}

where $v$, $I_p$ and $v_0$ are the wavelength vector (converted into velocity through Doppler effect), peak intensity and line center. \cite{Chae1998} defined $\sigma$ as Gaussian width. While the Gaussian width mentioned by \cite{Peter2010} is $\sqrt{2}\sigma$. In \cite{Tian2011c} we followed \cite{Peter2010} and used both the names of Gaussian width and 1/e width for $\sqrt{2}\sigma$. To avoid confusion, in the following we use the name exponential width (or 1/e width, also used by \cite{Peter2010}) instead of Gaussian width for $\sqrt{2}\sigma$. 

The technique of RB asymmetry analysis was first introduced by \cite{DePontieu2009} and it is based on a comparison of the two wings of the line profile at same velocity ranges. The line profile was first interpolated to a spectral resolution ten times greater than the original one, then the blue wing emission integrated over a narrow spectral range was subtracted from that over the same range in the red wing. The range of integration was then sequentially stepped outward from the line centroid to build an RB asymmetry profile (simply RB profile). In our previous work
\citep{DePontieu2009,DePontieu2010,DePontieu2011,McIntosh2009a,McIntosh2009b,McIntosh2011,Tian2011a,Martnez-Sykora2011}, we used the
single Gaussian fit to determine the line centroid and applied this technique to spectra in coronal holes, quiet Sun, and quiet ARs. \cite{Tian2011c} named this method RB$_{S}$ and they further developed two other methods RB$_{P}$ and RB$_{D}$,  which are basically the same as RB$_{S}$ except the determination of the line centroid. For RB$_{P}$, the spectral position corresponding to the peak intensity is used as the line centroid and the resulting RB profile is normalized to the peak intensity. For RB$_{D}$, the line center of the primary component, which is derived from the RB$_{P}$-guided double Gaussian fit, is used as the line centroid and the resulting RB profile is normalized to the peak intensity of the primary component. As pointed out by \cite{Tian2011c}, the RB$_{P}$ technique can resolve the blueshifted secondary component more accurately as compared to the originally defined RB$_{S}$ technique. Thus, here we apply the newly developed RB$_{P}$ and RB$_{D}$ techniques, as well as the RB$_{P}$-guided double Gaussian fit \citep[for details see][]{Tian2011c}, to the data in this paper.  

Figures~\pref{fig.1}-\pref{fig.6} show the spatial distributions of the peak intensity, velocity and exponential width derived from the single Gaussian fit, and the average RB$_{P}$ asymmetry in the velocity interval of 70-130~km~s$^{-1}$ for Fe~{\sc{xiii}}~202.04\AA{} or Fe~{\sc{xii}}~195.12\AA{} in the observations of six events. For the RB$_{P}$ asymmetry a negative/positive value indicates an enhancement of the blue/red wing. We prefer to use the Fe~{\sc{xiii}}~202.04\AA{} line to detect asymmetry since there is no identified blends in this strong line, although the pervasive presence of very weak redward asymmetries outside the dimming regions (in Figures~\pref{fig.1}-\pref{fig.3}) might suggest an unidentified weak blend at the red wing of the line profile. However, in Figures~\pref{fig.4}-\pref{fig.6} we present results for the Fe~{\sc{xii}}~195.12\AA{} line since the exposure time used in the associated observations is too short so that only the strong Fe~{\sc{xii}}~195.12\AA{} line has enough S/N to allow a reliable RB asymmetry analysis to individual profile. Since the blend Fe~{\sc{xii}}~195.18\AA{} sits at the red wing of Fe~{\sc{xii}}~195.12\AA{} \citep{Young2009}, any blueward asymmetries detected by our RB$_{P}$ technique are not caused by this identified blend.  

The pre-eruption parameters are presented in the first row of each figure. The scanned regions for all rasters are almost the same for almost every event. The only exception is the 2007 May 19 event shown in Figure~\pref{fig.2}, where we can clearly see that the observed region in the pre-eruption phase is about $50^{\prime\prime}$ smaller in solar Y, compared to that in the eruption phase. 

\section{Flows}

We found various types of flows in our observations. In the following we mainly investigate properties of three types of outflows associated with coronal dimmings and CME or EUV jet eruptions. 

\subsection{High-speed outflows in dimming regions}

\begin{figure*}
\centering {\includegraphics[width=0.98\textwidth]{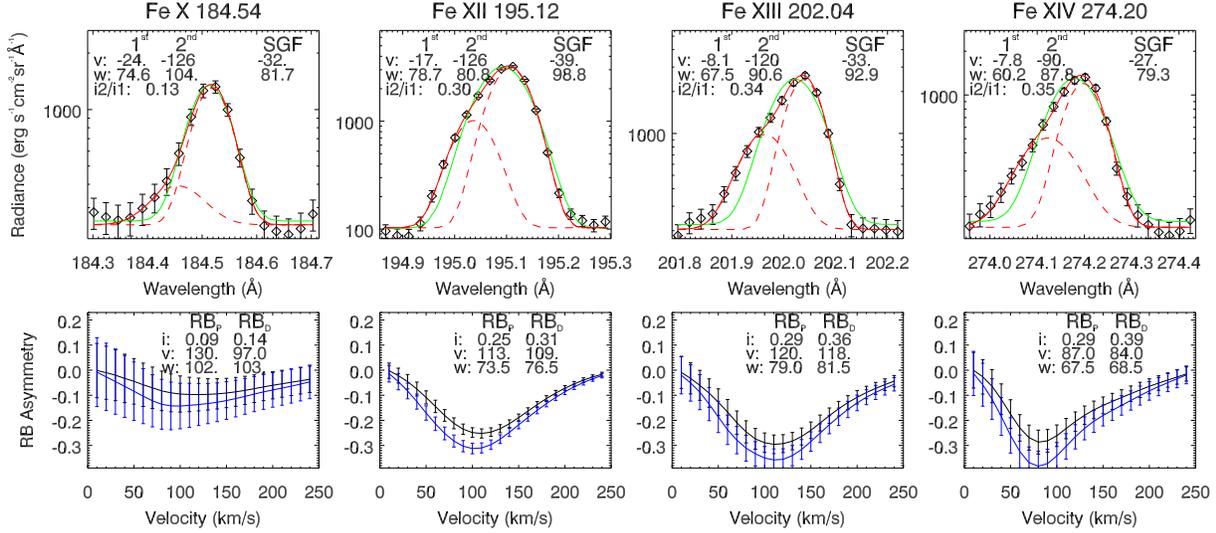}} \caption{ RB asymmetry profiles (bottom) of the Fe~{\sc{x}}~184.54\AA{}, Fe~{\sc{xii}}~195.12\AA{}, Fe~{\sc{xiii}}~202.04\AA{} and Fe~{\sc{xiv}}~274.20\AA{} line profiles (top) averaged over the square marked in Figure~\pref{fig.1}. Top: The observed spectra and measurement errors are shown as the diamonds and error bars, respectively. The green lines are single Gaussian fits. The two dashed red lines in each panel represent the two Gaussian components and the solid red line is the sum of the two components. The velocity (v) and exponential width (w) derived from the single (SGF) and double (1$^{st}$/2$^{nd}$ for the two components) Gaussian fits are shown in each panel. Also shown is the intensity ratio of the secondary component to the primary one (i2/i1). Bottom: the black and blue lines represent RB profiles for RB$_{P}$ and RB$_{D}$, respectively. Error bars indicate the errors propagated from the measurement errors. The peak relative intensity (i), velocity (v), and 1/e width (w) are shown in each panel. } \label{fig.7}
\end{figure*}

\begin{figure*}
\centering {\includegraphics[width=0.98\textwidth]{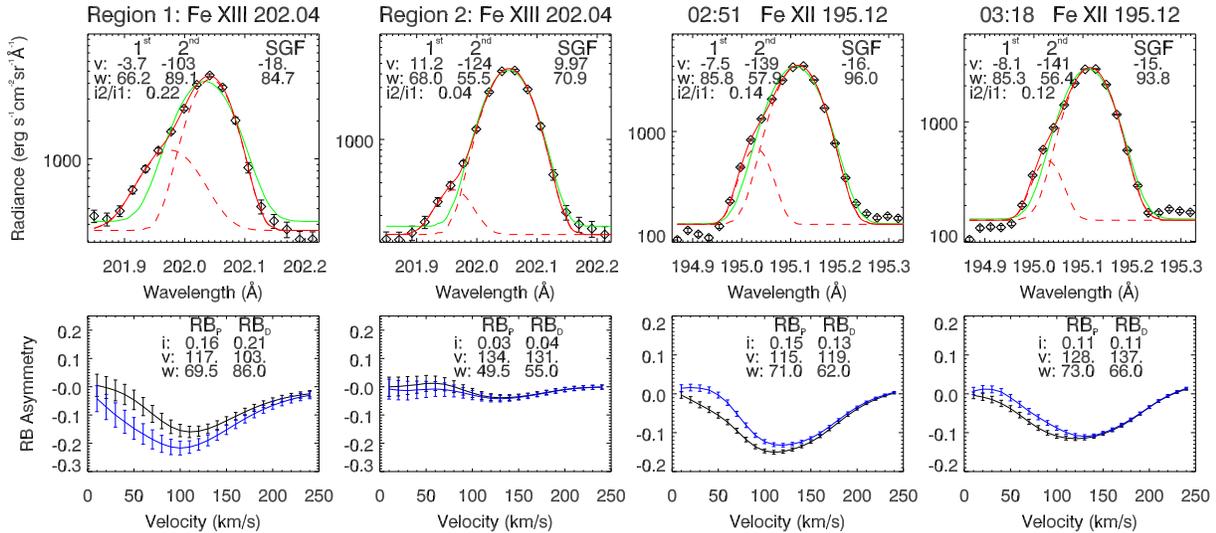}} \caption{ First \& second columns: RB asymmetry profiles (bottom) of Fe~{\sc{xiii}}~202.04\AA{} line profiles (top) averaged over regions 1 \& 2 marked in Figure~\pref{fig.2}. Third \& fourth columns: RB asymmetry profiles (bottom) of the averaged Fe~{\sc{xii}}~195.12\AA{} line profiles (top) in dimming regions at 02:51 and 03:18 as shown in Figure~\pref{fig.4}.  The line styles and denotations of parameters are the same as in Figure~\pref{fig.7}.} \label{fig.8}
\end{figure*}

Coronal dimmings are clearly seen from the intensity images presented in Figures~\pref{fig.1}-\pref{fig.4}. The dimming regions are characterized by a blueshift of 10-40~km~s$^{-1}$, a notable phenomenon in the Hinode era \citep{Harra2007,Jin2009,Attrill2010,Chen2010a,Harra2011a}. Enhancement of the line width in dimming regions has also been reported by \cite{McIntosh2009c}, \cite{Chen2010a} and \cite{Dolla2011} and it is very clear from Figures~\pref{fig.1}-\pref{fig.4}. The significant blueshift and enhanced line width are similar to those found at the weak-emission boundaries of ARs \citep[e.g.,][]{Marsch2004,Marsch2008,Harra2008,DelZanna2008,DelZanna2011,Doschek2007,Doschek2008,Tripathi2009,He2010a,Murray2010,Brooks2011,Warren2011,Bradshaw2011,Scott2011,Young2012,Baker2012,Hara2008,DePontieu2009,DePontieu2010,McIntosh2009a,McIntosh2009b,McIntosh2011,Peter2010,Bryans2010,Ugarte-Urra2011,Martnez-Sykora2011,Tian2011a,Tian2011c}. 

Figures~\pref{fig.1}-\pref{fig.4} also reveal a significant blueward asymmetry in dimming regions. We note that the blueward asymmetries on maps of RB$_{S}$ asymmetry for the 2006 Dec 14-15 observations, which were presented by \cite{McIntosh2010}, are not so prominent as those in the RB$_{P}$ asymmetry maps in our Figure~\pref{fig.1}. This is because of the underestimation of the degree of asymmetry by the RB$_{S}$ technique \citep{Tian2011c}. The presence of obvious blueward asymmetries suggests that the line profiles in dimming regions probably contain a highly blueshifted secondary component besides the primary component. Such a scenario is similar to that of the chromospheric network and AR edges \citep[e.g.,][]{Hara2008,DePontieu2009,DePontieu2010,McIntosh2009a,McIntosh2009b,McIntosh2011,Peter2010,Bryans2010,Tian2011a,Tian2011c,Ugarte-Urra2011,Martnez-Sykora2011}, and thus we can perform a similar analysis of the line profiles. For further analysis, we only selected those locations where the average RB$_{P}$ asymmetry in the velocity range of 70-130~km~s$^{-1}$ is smaller than -0.03 (obvious blueward asymmetry) and the signal to noise ratio of the profile (defined as the ratio of the peak and background intensities) is larger than 8. The RB$_{P}$-guided double Gaussian fit algorithm \citep[see details in][]{Tian2011c} was then applied to the profiles at these locations. After the double Gaussian fit, we took the spectral position of the primary component as the line centroid and calculated the RB$_{D}$ asymmetry profile.

In Figure~\pref{fig.7}~\&~\pref{fig.8}, we present several examples of the observed and fitted line profiles and the corresponding RB$_{P}$ and RB$_{D}$ asymmetry profiles in dimming regions. By comparing the observed profiles with the different fitting profiles, we can clearly see the better performance of the double Gaussian fits and the deviations of the observed profiles from the single Gaussian fits. The RB asymmetry profile is basically the difference between the emission of the two wings as a function of spectral distance (expressed in the velocity unit) from the line center. Here a negative value means that the blue wing is enhanced with respect to the red wing at a certain spectral distance. The relative intensity and velocity of the secondary component can be derived from the peak of the RB asymmetry profile. And the 1/$e$ width of the RB asymmetry profile is taken to approximate the width of the secondary component \citep[see details in][]{Tian2011c}. As discussed in \cite{Tian2011c}, the blend Si~{\sc{vii}}~274.18\AA{} should not have an important influence on the results of our RB asymmetry analysis and double Gaussian fit for Fe~{\sc{xiv}}~274.20\AA{} since the two lines are very close to each other and the maps of the Fe~{\sc{xiv}}~274.20\AA{} line parameters resemble those of the clean Fe~{\sc{xiv}}~264.78\AA{} line. Moreover, as mentioned in the following, the contribution of Si~{\sc{vii}}~274.18\AA{} to the total emission is at most 5.4\%. The Fe~{\sc{xii}}~195.12\AA{} line is blended with Fe~{\sc{xii}}~195.18\AA{}, which sits at the red wing of Fe~{\sc{xii}}~195.12\AA{} so that any enhancement on the blue wing of the line profile is not caused by this identified blend. Thus, the blueward asymmetries of the Fe~{\sc{xii}}~195.12\AA{} line profiles we observed here are real. The Fe~{\sc{x}}~184.54\AA{} line is not as strong as the other three lines and there is a weak Fe~{\sc{xi}}~184.41\AA{} line at the blue side \citep{Brown2008}, making it difficult to derive the real degree of profile asymmetry. However, the Fe~{\sc{xi}}~184.41\AA{} line is about 210~km~s$^{-1}$ away from the Fe~{\sc{x}}~184.54\AA{} line and in normal conditions the two lines show up as two distinct peaks. So we believe that the enhancement between the two lines are due to the Fe~{\sc{x}}~184.54\AA{} emission from a high-speed ($\sim$100~km~s$^{-1}$) outflow. As can be seen from Figure~\pref{fig.7}, we usually found that the blueward asymmetry is present at different coronal temperatures and that the velocity of the high-speed upflow does not show a dramatic change with temperature. However, we are aware that an accurate velocity determination is beyond the ability of the EIS instrument because of the large instrumental width and the complication by blends. 

The ridge of enhanced line width for the 2007 May 19 11:41-15:23 scan, as shown in Figure~\pref{fig.2}, was previously reported by \cite{Chen2010a}. They found that this ridge corresponded to the outer edge of the dimming region. In the Dopplergram, blue shift of $\sim$10~km~s$^{-1}$ seems to be present in the southern part of the ridge \citep{Chen2010a}. While the northern part shows net red shift, which seemed to be omitted by \cite{Chen2010a}. From the map of RB asymmetry, we can see that this ridge is also charactered by clear blueward asymmetry. Typical line profiles in both the southern and northern parts of the ridge are presented in Figure~\pref{fig.8}. Results of our RB asymmetry analysis indicate that the high-speed upflow is present along the whole ridge and that its speed might be $\sim$100~km~s$^{-1}$.

\begin{figure*}
\centering {\includegraphics[width=0.9\textwidth]{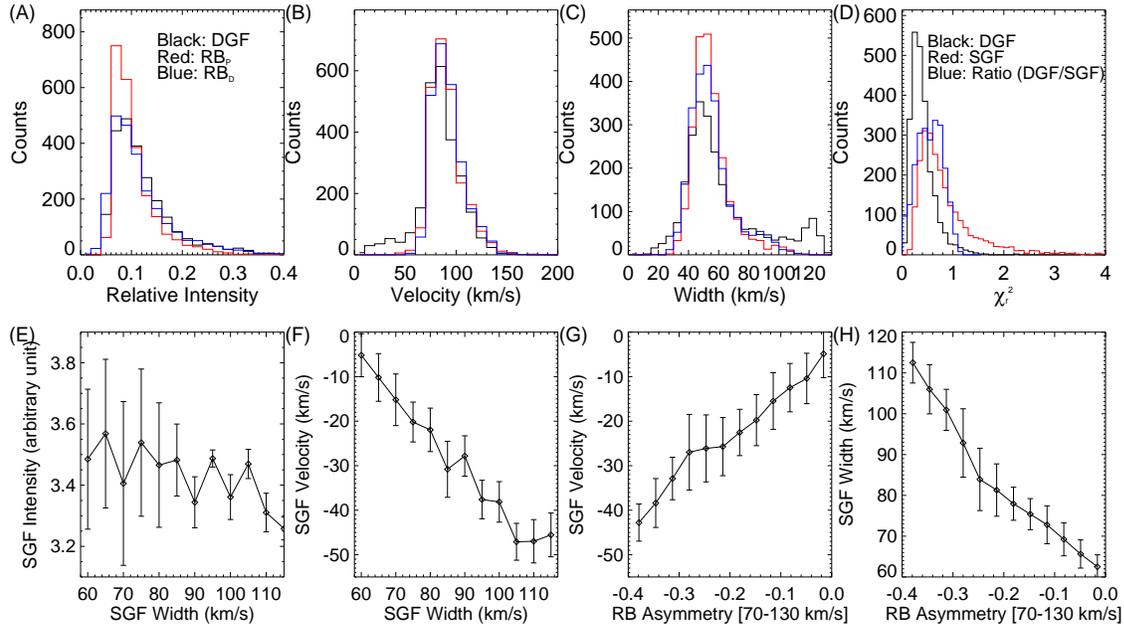}} \caption{ Histograms of the relative intensity (A), velocity (B), and
exponential width (C) of the secondary component, as derived from double Gaussian fit (black) and RB asymmetry analysis (red/blue for
RB$_{P}$/RB$_{D}$) for Fe~{\sc{xiii}}~202.04\AA{} in the 2006 Dec 15 01:15-03:30 observation. Panel (D) shows the histograms of the $\chi_{r}^{2}$ values of the single (red) and double (black) Gaussian fits, as well as the ratio of the two (blue). Panels (E)\&(F) show the relationship between the intensity/Doppler shift and the exponential width as derived from single Gaussian fit. Panels (G)\&(H) present the relationship between the Doppler shift/exponential width derived from single Gaussian fit and the average RB$_{P}$ asymmetry in the velocity interval of 70-130~km~s$^{-1}$. } \label{fig.9}
\end{figure*}

We derived the parameters (relative intensity, velocity, exponential width) of the high-speed secondary component by using the three methods (double Gaussian fit, RB$_{P}$ and RB$_{D}$) and in Figure~\pref{fig.9} we present the histograms of these three parameters derived from Fe~{\sc{xiii}}~202.04\AA{} line profiles in the 2006 Dec 15 01:15-03:30 observation (use profiles within the red contours shown in Figure~\pref{fig.1}). We can see that the relative intensity is usually around 10\% and can sometimes reach more than 30\%. The velocity is usually in the range of 50-150~km~s$^{-1}$ and its distribution peaks around 90~km~s$^{-1}$. The distribution of the exponential width peaks around 55~km~s$^{-1}$, which is comparable to the width of the primary component. The distributions of the $\chi_{r}^{2}$ for both the single and double Gaussian fits peak at values smaller than unity, which may result from the overestimation of the EIS measurement error \citep{Peter2010,Tian2011c}. However, from the $\chi_{r}^{2}$ ratio between the double and single Gaussian fit we can see that the double Gaussian fit does better than the single Gaussian fit for these asymmetric line profiles.

In Figure~\pref{fig.9} we also show the relationship between the intensity/Doppler shift and the exponential width as derived from single Gaussian fit, and the relationship between Doppler shift/exponential width derived from the single Gaussian fit (SGF) and the average RB$_{P}$ asymmetry in the velocity interval of 70-130~km~s$^{-1}$. There seems to be a weak anti-correlation between the SGF intensity and line width, which is consistent with previous result that the intensity and line width show negative correlations in loop footpoint regions \citep{Scott2011} although the correlation turns into positive when considering the whole AR \citep{Li2009}. Panel (F) reveals an obvious correlation between the SGF Doppler shift and line width. The calculated correlation coefficient is -0.59 if using all data points inside the red contours. Similar correlation was also found at AR boundaries \citep{Doschek2007,Doschek2008}. As proposed by \cite{Doschek2008}, this correlation may suggest that the profile is composed of multiple components. Indeed, we find striking correlations in panels (G)\&(H) of Figure~\pref{fig.9}, with a correlation coefficient of 0.67 in (G) and -0.58 in (H). Such correlations strongly suggest that the clear blue shift and enhanced line width in dimming regions are largely caused by the blueward asymmetries. The growth of Alfv\'en wave amplitude, as suggested by \cite{McIntosh2009c}, may be an additional reason for the enhancement of the line width. The fact may be that there is a faint high-speed upflow superimposed on a strong and almost stationary (or slightly shifted) background in the LOS direction. This scenario is also similar to the inhomogeneities of flow velocities along the LOS as proposed by \cite{Dolla2011} and would naturally produce blueward asymmetric line profiles. A single Gaussian fit to the total emission line profile would yield a blueshift and enhanced line width, as compared to the line profile of the background emission. When the relative intensity of the high-speed upflow component becomes larger, the blueward asymmetry becomes more obvious and we will obtain a larger blue shift and line width if applying a single Gaussian fit. The intrinsic assumption of a single Gaussian fit is that everything is moving at the same bulk speed, which is obviously not the case in coronal dimming regions. Thus, our analysis implies that previous results based on a single Gaussian fit can not reflect the real physical processes and thus may need to be reconsidered. First, the outflow speed in dimming regions is perhaps not around 20~km~s$^{-1}$\citep{Harra2007,Jin2009,Attrill2010,Chen2010a,Harra2011a}, but can easily reach $\sim$100~km~s$^{-1}$ in the lower corona. Second, the enhanced line width is not purely due to the increase of the Alfv\'en wave amplitude \citep{McIntosh2009c}, but is largely contributed by the superposition of different emission components. 

We note that the properties of these high-speed outflows are very similar to those we found previously in AR edges \citep{Tian2011c}. This similarity suggests that the outflows in both regions may result from a similar process, e.g., heating in the lower atmosphere \citep{DePontieu2009,McIntosh2009b,Hansteen2010,Song2011}. Magnetohydrodynamic simulations have shown that magnetic reconnection is an efficient mechanism to produce high-speed outflows (jets) in the lower solar atmosphere \citep[e.g.,][]{Ding2011,Roussev2001}. As the magnetic field lines opened up by CMEs, rapid multi-thermal upflows produced by both the pre-existing and CME-induced impulsive heating at the lower part of the erupted loops are guided by the field lines into the transiently opened corona. These outflows may serve as an important source of materials to refill the corona. From Figures~\pref{fig.1}\&\pref{fig.4} we can see that the blueward asymmetries were strongest within a few hours after the flare peak time, indicating that the high-speed outflows were strongest right after the erupted materials left the Sun. As the dimmings gradually recovered and the magnetic fields began to close down again, the outflows became weaker. Such a result is consistent with the finding of \cite{Miklenic2011} that the mass loss occurs mainly during the period of strongest CME acceleration.

Through joint imaging and spectroscopic observations of the corona, \cite{McIntosh2009a}, \cite{DePontieu2010} and \cite{Tian2011a,Tian2011c} have suggested that the secondary emission component found at AR edges is caused by high-speed repetitive upflows in the form of upward propagating disturbances (PDs) in EUV and X-ray imaging observations. Similarly, we think that the highly blueshifted component found in spectra of dimming regions should exhibit as PDs in imaging observations. The SDO/AIA observations, with a high S/N (especially in the 171\AA{} \& 193\AA{} passbands) and high cadence, might be able to reveal such weak PDs. The 2011 June 21 observations were done in the SDO era and from the associated movie (m4.mpeg) of Figure~\pref{fig.4} we can clearly see the evolution of the dimming boundary, which is likely to be associated with the successive disappearance of the "moss" \citep{McIntosh2007}. And, there seems to be weak outward PDs along legs of the opened coronal loops. However, both the LOS effect and the significantly reduced emission associated with these opened field lines make it difficult to study the PDs quantitatively. Part of these outflows may experience further acceleration at higher layers, overcome the gravity, and eventually become the solar wind stream along the transiently opened field lines, which may serve as an additional momentum source for the associated CME \citep{McIntosh2010}. We also noticed that the ascending post-flare loops revealed by the movie were clearly observed by EIS (last row of Figure~\pref{fig.4}), showing a blue shift of $\sim$5~km~s$^{-1}$.

\subsection{Temperature-dependent outflows}

\begin{figure*}
\centering {\includegraphics[width=0.98\textwidth]{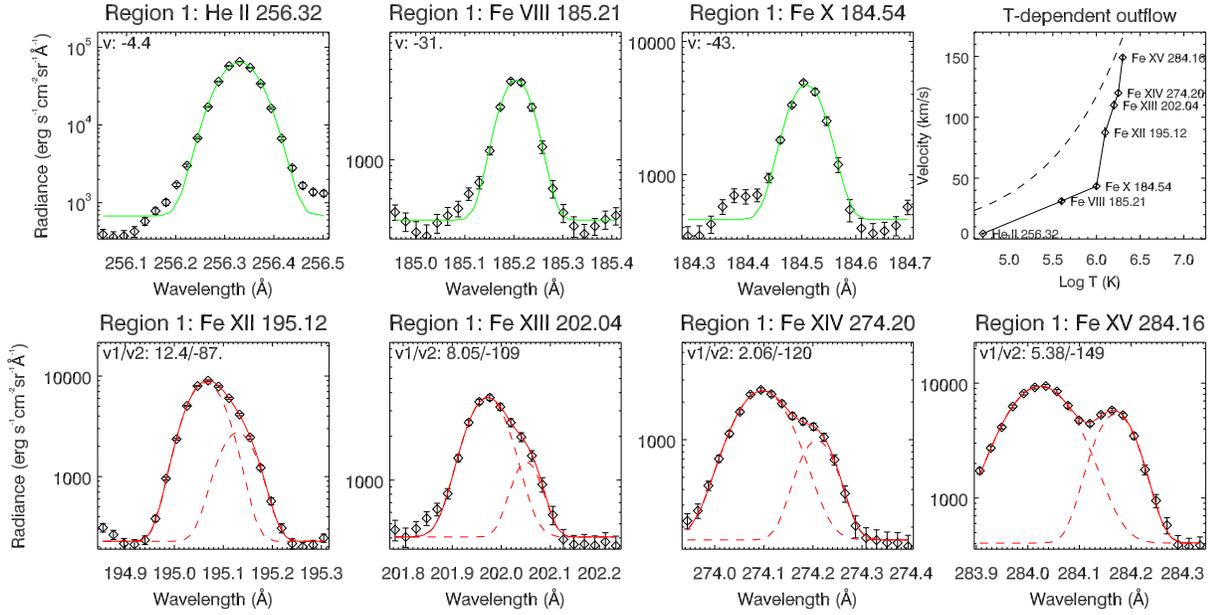}} \caption{ Line profiles averaged over region 1 marked in Figure~\pref{fig.3}. The diamonds, error bars, line styles and colors are the same as in Figure~\pref{fig.7}. The velocities (v) derived from the single (SGF) and double (v1 \& v2 for the two components) Gaussian fits are shown in each panel of line profile. Also shown is the temperature dependent outflow velocity and the adiabatic sound speed (dashed curve in the upper right panel). } \label{fig.10}
\end{figure*}

A temperature-dependent outflow was reported by \cite{Imada2007}, who found that the flow speed increases from $\sim$10~km~s$^{-1}$ at log ({\it T}/K)=4.9 to $\sim$150~km~s$^{-1}$ at log ({\it T}/K)=6.3 in a dimming region. However, we found that this temperature-dependent outflow is not in but immediately outside the deepest (darkest in the intensity image) dimming region. More interestingly, we found that our RB$_{P}$ technique can identify this temperature-dependent outflow. This event was located around (x=135$^{\prime\prime}$,y=-35$^{\prime\prime}$) in Figure~\pref{fig.3} and we can see that it is associated with a small patch of redward asymmetry. This is easy to understand since the outflow component is much stronger than the background emission component in Fe~{\sc{xiii}}~202.04\AA{}, as can be seen from Figure~\pref{fig.3}f of \cite{Imada2007}.

We further identified several temperature-dependent outflows in the 2006 Dec 13 01:12-05:41 observation. These temperature-dependent outflows are associated with the small patch of redward asymmetry around the location of (x=130$^{\prime\prime}$,y=-85$^{\prime\prime}$), (x=175$^{\prime\prime}$,y=-95$^{\prime\prime}$), (x=375$^{\prime\prime}$,y=-135$^{\prime\prime}$) and (x=135$^{\prime\prime}$,y=-35$^{\prime\prime}$), respectively. All of these temperature-dependent outflows are not in but immediately outside the deepest dimming region. As an example, Figure~\pref{fig.10} shows the line profiles of the temperature-dependent outflow around (x=175$^{\prime\prime}$,y=-95$^{\prime\prime}$), region 1 in Figure~\pref{fig.3}. A single Gaussian fit seems to be adequate to derive the outflow velocities for the emission lines formed at a temperature of log ({\it T}/K)$\leqslant$6.0. For emission lines formed at higher temperatures, we see clear indications of two well-separated components in the line profiles. Thus, we applied a double Gaussian fit to these line profiles and the Doppler shift of the highly blueshifted component (denoted as v2 in Figure~\pref{fig.10}) should represent the outflow velocity at the corresponding temperature. The small-velocity component, whose velocity is denoted as v1, is likely to be the nearly stationary background emission of the corresponding ion. Since we are mainly interested in the velocity of the highly blueshifted component and this component is often stronger than the background component, the outflow velocity derived from the double Gaussian fit should be highly reliable. We can see that the temperature variation of the outflow velocity shown in Figure~\pref{fig.10} is similar to Figure~\pref{fig.6} of \cite{Imada2007}. 

\begin{figure*}
\centering {\includegraphics[width=0.98\textwidth]{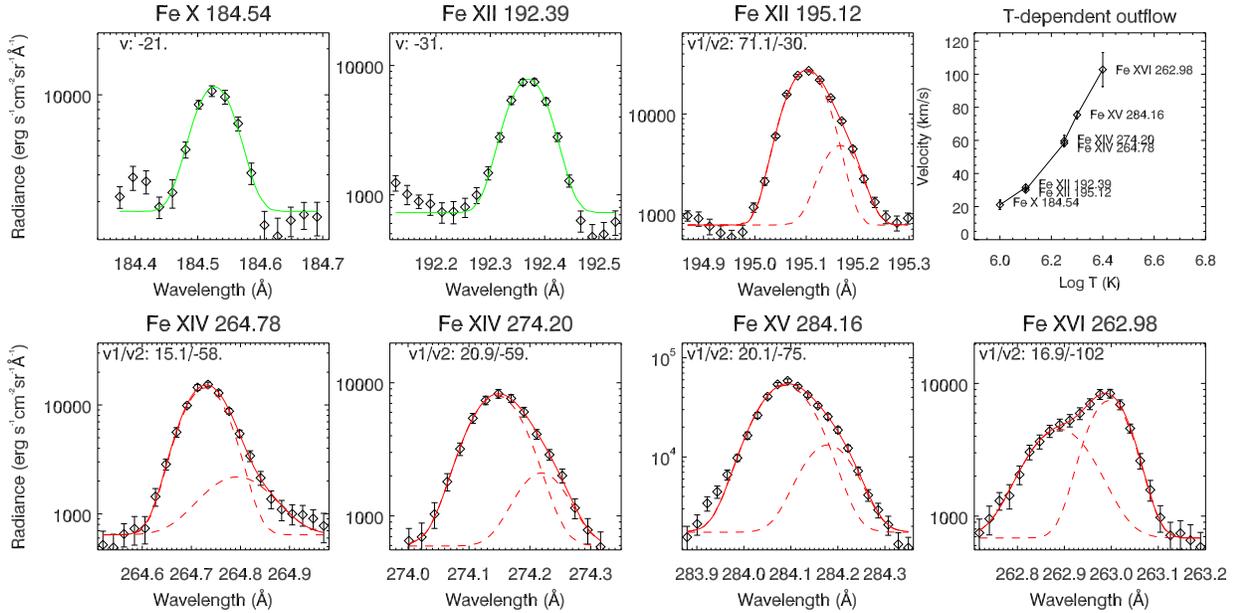}} \caption{Same as Figure~\pref{fig.10} but for the rectangular region marked in Figure~\pref{fig.4}. } \label{fig.11}
\end{figure*}

We also identified temperature-dependent outflows in the 2011 Jun 21 observation. Unfortunately, the exposure time used in this observation is only 9 seconds so that only the Fe~{\sc{xii}}~195.12\AA{} line can be used for asymmetry analysis without any temporal or spatial binning. From Figure~\pref{fig.4} we can see enhanced redward asymmetries surrounding the region of significant blueward asymmetry (dimming region). The blend of Fe~{\sc{xii}}~195.18\AA{} certainly contributes to these redward asymmetries. However, if we spatially bin line profiles of various lines at several adjacent pixels there, we start to see clear signatures of temperature-dependent outflows. Thus, both the blend of Fe~{\sc{xii}}~195.18\AA{} and the temperature-dependent outflows are causing these redward asymmetries. As an example, we present in Figure~\pref{fig.11} the line profiles averaged over the small rectangular region marked in Figure~\pref{fig.4}. Similar to Figure~\pref{fig.10}, single or double Gaussian fits are applied to these line profiles to obtain the outflow velocities at different temperatures. 

From Figures~\pref{fig.3}\&\pref{fig.4} we can also see the enhancement of the line width at locations where the temperature-dependent outflows are found. These enhanced line widths, as derived from single Gaussian fits, are actually caused at least partly by the superposition of the relatively weak background emission component and the strong outflow component for Fe~{\sc{xiii}}~202.04\AA{} and Fe~{\sc{xii}}~195.12\AA{}, as demonstrated in Figure~\pref{fig.10} and Figure~\pref{fig.11}, respectively. 

Note that there are several blends of the He~{\sc{ii}}~256.32\AA{} line. However, the He~{\sc{ii}}~256.32\AA{} line usually dominates and contributes more than 80\% of the total emission in disk observations \citep{Young2007}. The Fe~{\sc{viii}}~185.21\AA{} is blended with Ni~{\sc{xvi}}~185.23\AA{} but the blend should not have a large impact on the derived velocity, since the two lines are very close to each other and the latter is much weaker than the former \citep{Young2007}. The enhancement of the blue wing in the Fe~{\sc{xv}}~284.16\AA{} line profile, at around 283.95\AA{} in Figure~\pref{fig.11}, seems to be caused by the weak blend Al~{\sc{ix}}~284.03\AA{}\citep{Young2007} and should not impact the derived velocity of the very strong outflow component significantly. 

The fact that these temperature-dependent outflows are found outside the (deepest) dimming regions suggests that these outflows are different from the high-speed outflows we described in the previous section. The temperature-dependent nature of these outflows resembles that of gentle (as opposed to explosive) chromospheric evaporation flows. Gentle chromospheric evaporation can be driven by low-flux ($\le$10$^{10}$~ergs~cm$^{-2}$~s$^{-1}$) nonthermal electron beams in the flare impulsive phase \citep[e.g.,][]{Milligan2006b} or thermal conduction in the flare decay phase \citep[e.g.,][]{Antiochos1978,Berlicki2005}. The temperature-dependent outflows we present here are away from the flare sites so that they may not be directly related to the associated flares at first thought. However, we can not exclude the possibility that some magnetic field lines there are connected to the flare sites and that nonthermal electron beams or enhanced thermal conduction resulting from the flares cause the evaporation flows. For the 2006 Dec 12-13 event we may exclude the possibility of nonthermal electron beams since most temperature-dependent outflows were identified after the flare peak time. It is also possible that interactions between the opened field lines in the dimming region and the closed loops in the surrounding plage region produce low-flux nonthermal electrons or/and enhanced thermal conduction which will then generate the evaporation flows. 

The difference between the high-speed outflows in dimming regions and the temperature-dependent outflows immediately outside the (deepest) dimming regions can also be understood in the sense of driving force. The former are perhaps driven by magnetic reconnection in the chromosphere or TR. These multi-thermal outflows usually do not show obvious temperature dependence since the acceleration by magnetic force finishes at locations very close to the reconnection site \citep[e.g.,][]{Yokoyama1995}. As discussed above, the latter seem to be evaporation flows which are driven by pressure gradient force. Thus, the acceleration will continue as long as a pressure gradient exists \citep[e.g.,][]{Kamio2009,Shimojo2001,Judge2012}. The dramatic increase of the flow speed from log ({\it T}/K)$\leq$6.0 to log ({\it T}/K)$\ge$6.0, as shown in Figure~\pref{fig.10}, is similar to that of the event analyzed by \cite{Imada2007} and may be caused by a steep pressure (temperature) gradient at a certain height \citep{Imada2011}. 

\subsection{Highly blueshifted component representing the ejecta emission}

\begin{figure*}
\centering {\includegraphics[width=0.98\textwidth]{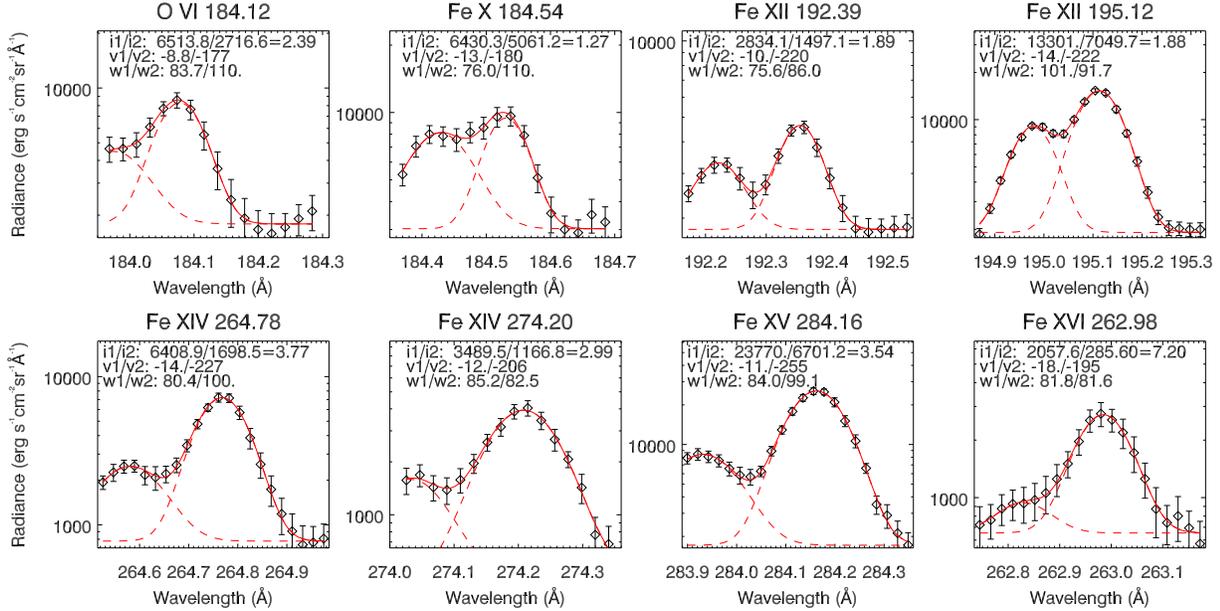}} \caption{ Line profiles averaged in locations marked in Figure~\pref{fig.5}. The diamonds, error bars, line styles and colors are the same as in Figure~\pref{fig.7}. The peak intensities (i1 \& i2), velocities (v1 \& v2) and exponential widths (w1 \& w2) of the two Gaussian components are shown in each panel. } \label{fig.12}
\end{figure*}

\begin{figure*}
\centering {\includegraphics[width=0.98\textwidth]{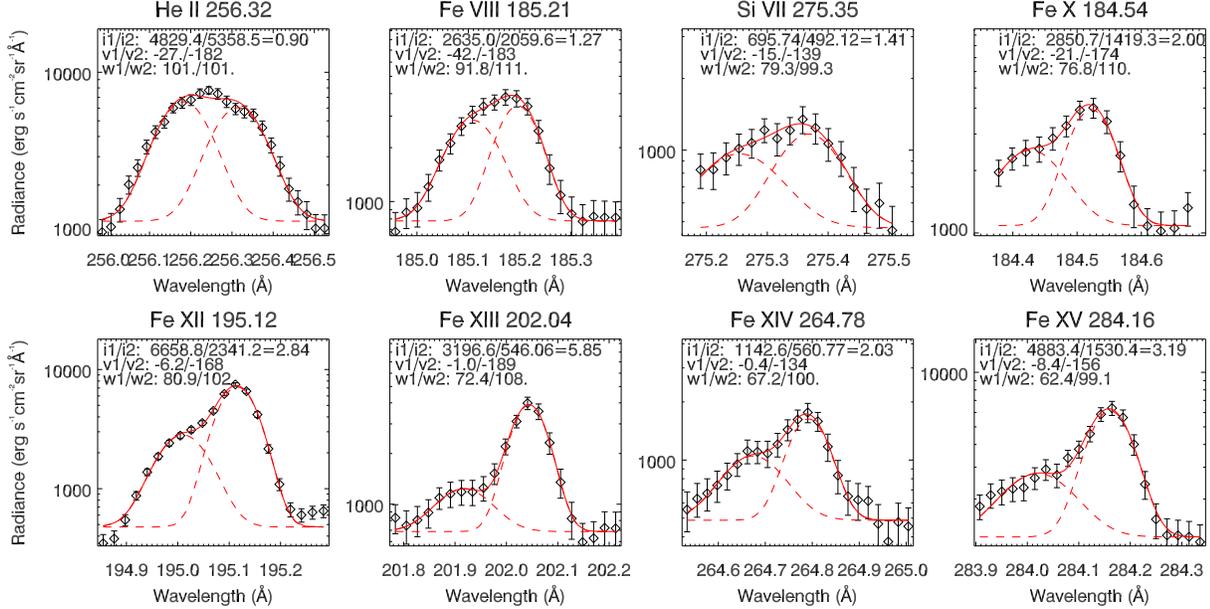}} \caption{Same as Figure~\pref{fig.12} but for the rectangular region marked in Figure~\pref{fig.6}. } \label{fig.13}
\end{figure*}

No pronounced dimming was recorded by EIS for the 2011 Feb 14 and 2007 Jun 5 observations. However, clear plasma ejections (CME or EUV jet) were observed and the associated line profiles clearly exhibit two well-separated components. The eruption of the CME loop on 2011 Feb 14 was clearly revealed in the AIA 304\AA{}, 171\AA{} \& 193\AA{} passbands (see m5.mpeg showing the evolution of the 193\AA{} passband, the green box indicates the FOV of EIS observation). For imaging observations of the 2007 Jun 5 jet, we refer to \cite{YangL2011}. Figures~\pref{fig.12}\&\pref{fig.13} show an example of line profiles associated with the CME ejecta and the EUV jet, respectively. It is clear that emission lines formed at coronal and TR temperatures clearly exhibit two well-separated components, an almost stationary component accounting for the background emission and a highly blueshifted component representing emission from the erupting material. The Doppler velocities of the two components can be easily calculated through a double Gaussian fit. 

From Figure~\pref{fig.12} we can see that the highly blueshifted component has a velocity of $\sim$220~km~s$^{-1}$ and the velocity does not change significantly with temperature. This velocity should represent the LOS velocity of the CME ejecta (expanded loops) at 19:29 on 2011 Feb 14. The POS component of the ejecta velocity was estimated to be $\sim$200~km~s$^{-1}$ from simultaneous AIA 193\AA{} observations. Combining the two velocity components, we can calculate the real speed of the ejecta at 19:29, which is $\sim$300~km~s$^{-1}$. Unfortunately, the fast EIS scans  only focused on the same region so that the oblique propagation of the CME ejecta was not fully tracked. But in principle one should be able to track the complete evolution (not only the POS component) of CMEs by using simultaneous imaging and spectroscopic observations. We noticed that clear line splittings interpreted as filament or plasmoid eruptions at similar speeds were previously reported by \cite{Harra2003} and \cite{Asai2008}.

From Figure~\pref{fig.13} we can see that the blueshifted component, which is apparently associated with the EUV jet, has a speed of $\sim$170~km~s$^{-1}$. The fact that we do not see an obvious temperature dependence of the flow speed suggests that this EUV jet is produced by reconnection instead of evaporation \citep{Kamio2009}. Since the POS component of the jet speed is $\sim$145~km~s$^{-1}$ \citep{YangL2011}, the real speed is calculated to be $\sim$223~km~s$^{-1}$. From Figure~\pref{fig.13} we can also see that the intensity ratio of the nearly stationary component and the highly blueshifted component (i1/i2) increases from 0.90 at log ({\it T}/K)=4.7 to 5.85 at log ({\it T}/K)=6.2. Such an increase should be directly related to the difference in the temperature distribution (DEM, discussed below) of the background emission and the jet emission. It is likely that the decrease of blue shift with temperature, as derived by \cite{YangL2011} from a single Gaussian fit, is in fact caused by this increase of intensity ratio and thus can not reflect the real physical process. In addition, the very large non-thermal velocities ($\sim$100-400~km~s$^{-1}$) reported by \cite{Kim2007} and \cite{YangL2011} through single Gaussian fits are also likely to be caused by the effect of the superposition of the two (background and jet) emission components. 

Another feature that is worth noting is the inverted Y-shape structure at one footpoint of the erupted loop (the fifth row of Figure~\pref{fig.5}) and the base of the jet (the second row of Figure~\pref{fig.6}). A net red shift and enhanced line width are found at the base of the inverted Y-shape structure in each case. Red shifts have been previously reported at the base of a polar jet and an AR jet by \cite{Kamio2007} and \cite{Chifor2008a}, respectively. It is likely that they are caused by the downward propagating reconnection outflows which collide with and compress the underlying loops \citep{Yokoyama1995}. And the enhanced line widths perhaps result at least partly from the flow inhomogeneities in this process. It seems that significant redward asymmetries are also found at the bases of the inverted Y-shape structures. But they are complicated by the blend Fe~{\sc{xii}}~195.18\AA{}. However, by averaging profiles of some weaker but clean lines over several spatial pixels at the bases of the inverted Y-shape structures, we do see signatures of redward asymmetries (not shown here) which may be caused by the downward flows. 

\begin{figure*}
\centering {\includegraphics[width=0.98\textwidth]{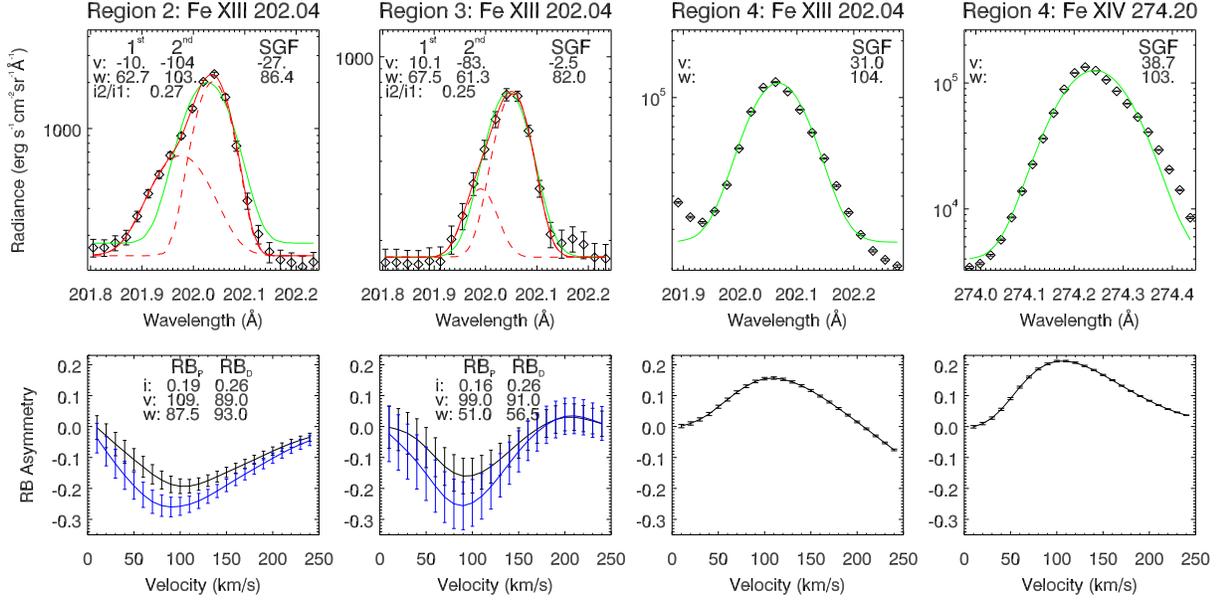}} \caption{Similar to Figure~\pref{fig.7} but for regions 2-4 marked in Figure~\pref{fig.3}. For the averaged profiles in region 4 only the single Gaussian fits and the RB$_{P}$ asymmetry profiles are plotted.} \label{fig.14}
\end{figure*}

At the end of this section, we would like to discuss various types of flows found in the 2006 Dec 12-13 observations. For detailed descriptions of this event, we refer to \cite{Kubo2007}, \cite{Zhang2007}, \cite{Asai2008}, \cite{Jing2008} and \cite{Fan2011}. From Figure~\pref{fig.3} we can see that before eruption (19:07-23:46) the loop footpoint regions are characterized by clear blue shift, enhanced line width and obvious blueward asymmetry. Such results indicate the presence of a weak high-speed upflow superimposed on the nearly stationary background \citep[e.g.,][]{DePontieu2009,McIntosh2009a,Tian2011a,Tian2011c,Martnez-Sykora2011}. During eruption (01:12-05:41) there was a clear expansion of the regions with clear blue shift, enhanced line width and obvious blueward asymmetry. Line profiles in the newly formed dimming regions (e.g., profile of Region 3 shown in Figure~\pref{fig.14}) are obviously blueward asymmetric and thus are similar to those of loop footpoint regions, indicating the presence of rapid upflows along the opened field lines as discussed in Section 3.1. The elongated blueward-asymmetry feature was observed in the flare impulsive phase and it was not located in the dimming region. The line profiles there (e.g., profile of Region 2 shown in Figure~\pref{fig.14}) clearly reveal two components and the high-speed component is likely to be associated with the initial removal of the magnetic loop. From Figure~\pref{fig.3} we can also see patches of significant redward asymmetries around the flare site (around x=330$^{\prime\prime}$,y=-80$^{\prime\prime}$). Figure~\pref{fig.14} shows profiles of two lines in a small region (Region 4 marked in Figure~\pref{fig.3}) and the redward asymmetries are clearly revealed in both the single Gaussian fits and RB$_{P}$ asymmetry profiles. Such asymmetries clearly indicate that the enhanced nonthermal broadening is caused by the superposition of flows, i.e., turbulence \citep{Milligan2011}. We found a net red shift of $\sim$30~km~s$^{-1}$ for almost all strong lines used in this observation. This multi-thermal downward motion is perhaps driven by both the cooling of the flare plasma and the overpressure of the flare plasma relative to the underlying atmosphere.

\section{Plasma diagnostics}

\subsection{Dimmings}

Using CDS data, \cite{Harrison2000} and \cite{Harrison2003} made first efforts to diagnose the electron densities and calculate the mass losses of dimming regions observed at the limb. They concluded that the reduced emission in dimming regions is an effect of mass loss rather than temperature change. They also mentioned the importance of calculating mass losses of on-disk dimming regions in the context of space weather forecast. Based on static solar atmosphere models, \cite{Jin2009} also tried to calculate the mass losses of two dimming regions by using EIS observations. 

\begin{figure*}
\centering {\includegraphics[width=0.98\textwidth]{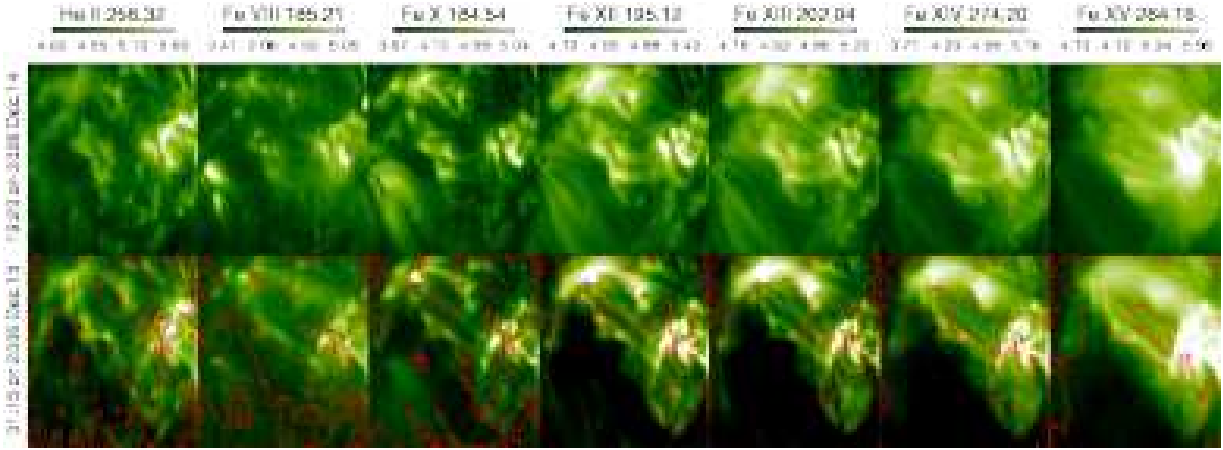}} \caption{Intensity maps of different emission lines for the two scans starting at 19:20 on 2006 Dec 14 and 01:15 on 2006 Dec 15. The red contours outline regions where the intensity was significantly reduced ($\geqslant$20\%). } \label{fig.15}
\end{figure*}

\begin{figure}
\centering {\includegraphics[width=0.45\textwidth]{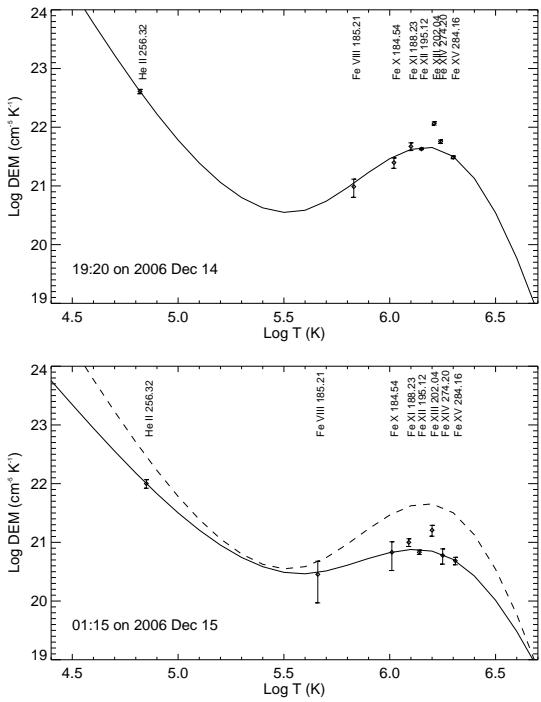}} \caption{DEM curves for the two scans starting at 19:20 on 2006 Dec 14 (upper panel, before eruption) and 01:15 on 2006 Dec 15 (lower panel, dimming). The DEM curve before the eruption is overplotted as the dashed line in the lower panel. } \label{fig.16}
\end{figure}

Here we extend these previous investigations and make efforts to diagnose the electron density, temperature and mass loss for the dimming regions we study in this paper. By comparing the intensity images of various emission lines before (19:20-21:35) and after (01:15-03:30) the eruption for the 2006 Dec 14-15 event (Figure~\pref{fig.15}), we can clearly see the occurrence of dimming at all temperatures. We then averaged line profiles over the regions where the intensity was reduced more than 20\% and calculated intensities of different lines before and after the eruption. By using the routine {\it chianti\_dem.pro} \citep[also used by][]{Lee2011} available in SolarSoft and assuming a constant pressure of 10$^{16}$~cm$^{-3}$~K, we obtained the differential emission measure (DEM) curves at the pre-eruption phase and of the dimming region. Here a double Gaussian fit was applied to the Fe~{\sc{xi}}~188.23\AA{} line profiles to derive the input line intensities since Fe~{\sc{xi}}~188.23\AA{} is partly blended with the strong Fe~{\sc{xi}}~188.30\AA{} line \citep{Young2007}. While we applied single Gaussian fits to the profiles of other selected lines. This means that the blends to Fe~{\sc{xii}}~195.12\AA{} and Fe~{\sc{xiv}}~274.20\AA{} were simply neglected since their contribution to the total emission is of the order of 5\% or less in outer parts (low density) of ARs \citep[see below and][]{Young2009}. The influence of the blend Al~{\sc{ix}}~284.03\AA{} is also negligible since it is at the far wing of the Fe~{\sc{xv}}~284.16\AA{} line and is very weak in AR conditions. The He~{\sc{ii}}~256.32\AA{} line is blended with several other higher-temperature lines (Fe~{\sc{xiii}}~256.42\AA{}, Fe~{\sc{xii}}~256.41\AA{}, Si~{\sc{x}}~256.37\AA{}). We included this line for our DEM analysis since it is the only strong EIS line formed in the lower TR and it contributes more than 80\% of the total emission in disk observations \citep{Young2007}. Moreover, the blends are all sitting at the red wing of the He~{\sc{ii}}~256.32\AA{} line profile and their spectral distances from the He~{\sc{ii}}~256.32\AA{} line center are $\sim$60-120~km~s$^{-1}$. We note that in such a case our single Gaussian fit algorithm mainly fits the core and blue wing of the line profile, which is primarily the emission of He~{\sc{ii}}~256.32\AA{} rather than the blends. Note that the weak high-speed outflow was not considered in the DEM and the following density diagnostics since it only contributes a few percent to the integrated intensity of the average line profiles. We can see from Figure~\pref{fig.16} that the main difference is the reduced emission at high temperatures (log ({\it T}/K)=6.1-6.3) in the dimming region. This result seems to suggest that in this event a significant portion of the cool TR materials did not escape when the magnetic field lines opened up. Similar result has also been obtained by \cite{Robbrecht2010} based on EUV imaging observations. We have to mention that the lower-temperature part of the DEM curves is less constrained due to the lack of many cool lines in the EIS observation. Observations of IRIS, which will be launched in 2012, are thus crucial since several strong cool lines are included in its spectrum.   

\begin{figure}
\centering {\includegraphics[width=0.45\textwidth]{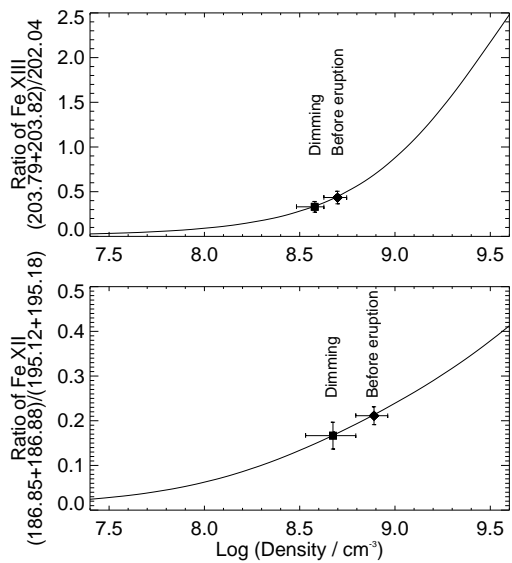}} \caption{ Relationship between electron density and line ratio, as derived from the CHIANTI database. The measured values before the eruption (15:11 on 2006 Dec 14) and during dimming (10:29 on 2006 Dec 15) are indicated by the solid diamonds and squares respectively. The vertical and horizontal bars indicate the standard deviations of the intensity ratios and uncertainties of the calculated densities, respectively. } \label{fig.17}
\end{figure}

\begin{table*}[]
\caption[]{Mass losses estimated from different methods and outflow densities for three dimming regions} \label{table2}
\begin{center}
\begin{tabular}{p{0.6cm} | p{3.8cm} | p{1.8cm} | p{1.8cm} | p{1.5cm} | p{2.5cm}}
\hline Obs. ID & Scanning Period & 1$^{st}$ Method (g) & 2$^{nd}$ Method (g) & CME mass (g) & Outflow density / log ({\it N}$_{e}$/cm$^{-3}$) \\
\hline 
1 & 2006 Dec 14 19:20-21:35 & 1.9$\times10^{15}$ & 8.0$\times10^{14}$ & 3.6$\times10^{15}$ & 7.0 \\
\hline   
3 & 2006 Dec 13 01:12-05:41 & 4.1$\times10^{15}$ & 1.4$\times10^{15}$ & 7.0$\times10^{15}$ & 7.1 \\
\hline
4 & 2011 Jun 21 02:51-02:56 & 1.1$\times10^{15}$ &  5.0$\times10^{14}$ &  & 6.8 \\
\hline
\end{tabular}
\end{center}
\end{table*}

There are several density sensitive line pairs used in the first (15:11-16:01) and last scans (10:29-11:19) of the 2006 Dec 14-15 observation. Thus, we can compare the densities of the dimming region with pre-eruption densities. Line pairs Fe~{\sc{xii}}~186.88\AA{}~\&~195.12\AA{} and Fe~{\sc{xiii}}~203.82\AA{}~\&~202.04\AA{} were chosen for the density diagnostics. Fe~{\sc{xii}}~186.88 is actually a self blends of two Fe~{\sc{xii}} lines (186.85\AA{}, 186.88\AA{}) and the Fe~{\sc{xii}}~195.12\AA{} line is blended with Fe~{\sc{xii}}~195.18\AA{} \citep{Young2007}. The S~{\sc{xi}}~186.84\AA{} line usually contributes no more than 5\% of the Fe~{\sc{xii}}~186.88\AA{} feature and thus was ignored in our calculation \citep{Chifor2008a}. Fe~{\sc{xiii}}~203.82\AA{} line is a self blends of two Fe~{\sc{xiii}} lines (203.79\AA{}, 203.82\AA{}). We calculated the line ratios and the associated standard deviations in the obvious dimming region (defined as the region with an intensity reduction larger than 20\%) of the last scan and in the corresponding region of the first scan. The theoretical relationships between the line ratios and densities, as extracted from CHIANTI database \citep{Dere1997,Landi2006}, are presented in Figure~\pref{fig.17}. The measured values before the eruption and during dimming are indicated by the solid diamonds and squares respectively. We can see that the average density changes from log ({\it N}$_{e}$/cm$^{-3}$)=8.89 to log ({\it N}$_{e}$/cm$^{-3}$)=8.67 at log ({\it T}/K)=6.1, and from log ({\it N}$_{e}$/cm$^{-3}$)=8.70 to log ({\it N}$_{e}$/cm$^{-3}$)=8.58 at log ({\it T}/K)=6.2. This density decrease, together with the fact that the dimming is seen at all temperatures, strongly suggests that the dimming is an effect of density decrease rather than temperature change. 

Following \cite{Harrison2000} and \cite{Jin2009}, we have attempted to estimate the mass losses in several well-observed dimming regions by using two different methods. The first method is just to multiply the density change and the emission volume, which is similar to the Si~{\sc{x}} method used by \cite{Harrison2003}. The calculation process can be expressed as following:

\begin{equation}
\emph{$M=\delta NSLm_{p}$}\label{equation2},
\end{equation}

where $M$, $\delta N$, $S$, $L$ and $m_{p}$ represent the total mass loss, change of the number density, area of the dimming region, depth of the dimming region and proton mass, respectively. The density change $\delta N$ can be calculated from the line pair Fe~{\sc{xii}}~186.88\AA{}~\&~195.12\AA{}, or estimated from the intensity change of Fe~{\sc{xii}}~195.12\AA{} if the other line was not used in the scan of dimming. The use of the Fe~{\sc{xii}}~195.12\AA{} line seems to be reasonable since from Figure~\pref{fig.16} we can see that the most significant decrease as well as the DEM peak occur around log ({\it T}/K)=6.1. The dimming area $S$ is defined as the total area where the Fe~{\sc{xii}}~195.12\AA{} intensity drops more than 20\%, multiplied by the ratio of the total dimming area in simultaneous full-disk coronal images (EIT or AIA observations) and the dimming area observed by EIS. Assuming that the emission volume is as deep as it is wide, the depth of the dimming region can be calculated as $\sqrt{S}$. As mentioned by \cite{Harrison2003}, the mass calculated from this method should be considered to be a reasonable figure for a comparison with the mass of the associated CME, although the Fe~{\sc{xii}}~186.88\AA{}~\&~195.12\AA{} line pair can only detect density changes at a temperature of around log ({\it T}/K)=6.1. 


The second method is similar to that used by \cite{Jin2009}. We take the emission heights of TR lines from the VAL3C model \citep{Vernazza1981} and coronal lines from \cite{Mariska1978}, then calculate the densities at these heights from an empirical relationship between height and density \citep{Cox2000,Jin2009}. Any density sensitive line pairs available in EIS observations are then used to derived the densities, which are then compared with and scale the model densities. Density changes at different heights can then be derived from intensity changes of corresponding EIS lines. The total mass loss is then expressed by

\begin{equation}
\emph{$M=\sum \delta N(h_{i})S(h_{i})\delta h_{i}m_{p}$}\label{equation3},
\end{equation}

where $\delta N(h_{i})$ and $S(h_{i})$ are the density change and dimming area respectively at the emission height of the {\it i}th line.

The mass losses calculated from the two methods for several well-observed dimming events are listed in Table~\pref{table2}. It can be seen that the mass losses estimated from different methods are 20\%-60\% of the CME mass calculated from LASCO white light data \citep{Jin2009}. Such results indicate that a significant part of the CME mass originates from the dimming region. Thus, in principle the values of mass loss estimated from spectroscopic observations can be used to guide our identifications of CME/ICME sources. We have to mention that no strong TR lines were used in the 2011 Jun 21 observation so that the mass loss derived from the second method is likely to be underestimated. We think that the difference of the DEM curves before and during the dimming, as shown in Figure~\pref{fig.16}, can in principle be used to derive the mass loss of the dimming region. However, it seems that we still have to make several assumptions for the dimming depth as well as the changes of density and temperature gradient, which are hard to evaluate. Thus, we do not make an effort in this direction and leave it open for future investigations.

 \cite{Jin2009} made an effort to calculate the mass flux for the outflows in dimming regions and concluded that the total ejected mass is about one order of magnitude larger than the CME mass. The flow velocities they used were derived from a single Gaussian fit to line profiles in dimming regions and they are of the order of $\sim$20~km~s$^{-1}$. The density values were taken from static solar atmosphere models. If we assume that the mass refilling the corona comes from the high-speed outflows, we should expect to see an equivalence of the total mass supplied by the outflow and the mass loss in the corresponding dimming region. If we simply estimate the total mass supplied by the outflow as the product of mass flux density, area ($S$) and duration ($t$) of the dimming, we can have the following relationship: 
 
\begin{equation}
\emph{$\delta NSL=nvSt$}\label{equation4},
\end{equation} 

From EIT or AIA observations, we roughly estimated the duration of significant dimming as 14, 14 and 10 hours for the 2006 Dec 14, 2006 Dec 13 and 2011 Jun 21 events, respectively. If we take a value of 100~km~s$^{-1}$ for the speed ($v$), the density ($n$) of the outflow can then be calculated from Equation~\pref{equation4}. The calculated densities of the outflows are listed in Table~\pref{table2}. These values are about two orders of magnitude smaller than the pre-eruption densities at log ({\it T}/K)=6.1 and should only be regarded as the lower limits since some of the outflows may overcome the gravity and become part of the solar wind. The lower limit of the mass flux density associated with these high-speed outflows is thus estimated to be about 1.67$\times$10$^{-10}$ g~cm$^{-2}$~s$^{-1}$ if using a density of log ({\it N}$_{e}$/cm$^{-3}$)=7.0. While the average outflow mass flux density of the dimming region in the 01:15-03:30 scan on 2006 Dec 14 is estimated to be about 1.0$\times$10$^{-9}$ g~cm$^{-2}$~s$^{-1}$ if using results of single Gaussian fit, i.e., an average velocity of 12.5~km~s$^{-1}$ and a density of log ({\it N}$_{e}$/cm$^{-3}$)=8.67. These values are within one order of magnitude of the mass flux density of type-I spicules \citep[$\sim$1.67$\times$10$^{-9}$ g~cm$^{-2}$~s$^{-1}$,][]{Pneuman1978}, type-II spicules \citep[$\sim$1.5$\times$10$^{-9}$ g~cm$^{-2}$~s$^{-1}$,][]{DePontieu2011}, coronal rains \citep[$\sim$1.14$\times$10$^{-9}$ g~cm$^{-2}$~s$^{-1}$,][]{Antolin2012} and outflows in the quiet-Sun network \citep[$\sim$1.6$\times$10$^{-9}$ g~cm$^{-2}$~s$^{-1}$,][]{Tian2009}.

We have to point out that the calculations of the mass flux of the high-speed outflows are based on the assumption that there are only two emission components and that the primary component is at rest. However, in fact we can not tell whether the primary component is really at rest or moving upward with a small velocity (e.g., $\sim$10~km~s$^{-1}$) since EIS does not allow an absolute wavelength calibration. Moreover, due to the large instrumental width we can not rule out the possibility of more than two components with each slightly Doppler-shifted with respect to each other \citep[e.g.,][]{Doschek2008}. 

\subsection{Erupted CME loop and EUV jet}

\begin{figure}
\centering {\includegraphics[width=0.45\textwidth]{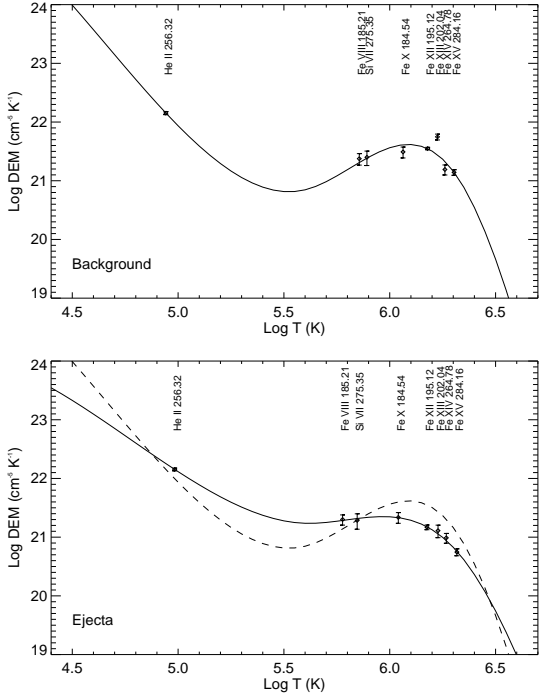}} \caption{DEM curves of the background (upper panel, stationary component) and ejecta (lower panel, highly blue-shifted component) for the rectangular region marked in Figure~\pref{fig.6}. The DEM curve of the background is overplotted as the dashed line in the lower panel. } \label{fig.18}
\end{figure}

In the case of erupted CME loops or EUV jets, since we can often unambiguously separate the ejecta emission component from the background emission component, e.g., Figures~\pref{fig.12}\&\pref{fig.13}, in principle we should be able to diagnose the plasma properties separately for each of the two components. Using the fitting results shown in Figure~\pref{fig.13}, we calculated the DEM curves of the two components. From Figure~\pref{fig.18} we can see that the ejecta (EUV jet) has more emission around log ({\it T}/K)=5.5, as compared to the background. We can also see that the emitting materials of the EUV jet are almost equally distributed over the temperature range of log ({\it T}/K)=5.4-6.1. We have to mention that there was only one line with a formation temperature of log ({\it T}/K)$\leq$5.5 (He~{\sc{ii}}~256.32\AA{}) in our EIS observations so that the low-temperature part of the DEM is not well constrained. Future joint observations of EIS and IRIS are thus highly desired. Unfortunately, there are no very cool lines (with a formation temperature comparable to that of He~{\sc{ii}}~256.32\AA{}) in the 2011 Feb 14 observation so that we could not perform a reliable DEM analysis (with a wide temperature coverage) for the CME ejecta in this observation.

\begin{figure}
\centering {\includegraphics[width=0.45\textwidth]{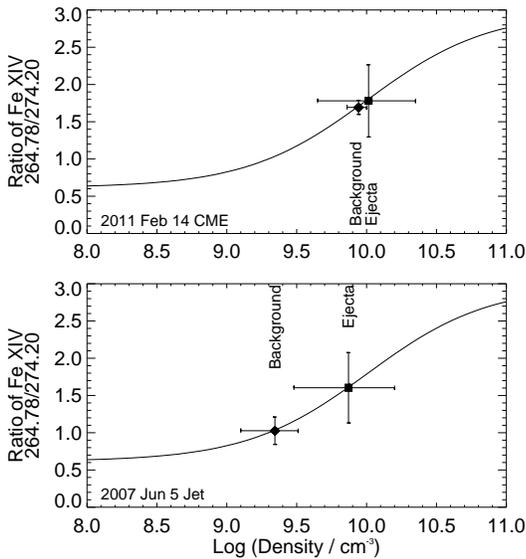}} \caption{Relationship between electron density and line ratio, as derived from the CHIANTI database. The measured values for the emission of the background and ejecta are indicated by the solid diamonds and squares respectively. The error bars indicate uncertainties of the ratios and densities as calculated from the 1-$\sigma$ uncertainties of the double Gaussian fit intensities. Upper: the 2011 Feb 14 CME; Lower: the 2007 Jun 5 jet.} 
\label{fig.19}
\end{figure}

The density sensitive line pair Fe~{\sc{xiv}}~264.78\AA{}~\&~274.20\AA{} were included in both the 2007 Jun 5 and 2011 Feb 14 observations. The Fe~{\sc{xiv}}~274.20\AA{} line is blended with Si~{\sc{vii}}~274.18\AA{}. The intensity of Si~{\sc{vii}}~274.18\AA{} can be estimated from Si~{\sc{vii}}~275.35\AA{} since the ratio of the two is at most 0.25 \citep{Young2007}. We used another observation which included both the Si~{\sc{vii}}~275.35\AA{} and Fe~{\sc{xiv}}~274.20\AA{} lines \citep{Tian2011c} and estimated that the contribution of Si~{\sc{vii}}~274.18\AA{} to the total emission is at most 5.4\%. Thus, we simply ignored this blend. The theoretical relationship between the line ratio and density, as well as the measured values for the two components, are shown in Figure~\pref{fig.19}. We can see that the measured values are in the density sensitive part of the theoretical relationship. The ejecta, which is an EUV jet in the 2007 Jun 5 observation and an erupted loop in the 2011 Feb 14 observation, seems to have a larger density compared to the background emission. The measured densities are log ({\it N}$_{e}$/cm$^{-3}$)=9.89 for the EUV jet and log ({\it N}$_{e}$/cm$^{-3}$)=10.01 for the erupted CME loop. However, the uncertainties of the line ratios for the ejecta are very large. This is largely due to the fact that the ejecta component in the Fe~{\sc{xiv}}~274.20\AA{} line profile was too close to the edge of the spectral window so that it was only partly resolved. The measured densities of the background components are consistent with (in the jet case) or about two times larger than (in the CME case) the normal AR densities at similar temperatures \citep{Tripathi2008}. To the best of our knowledge, this is the first time that the electron densities of EUV jets and erupted CME loops are measured through an unambiguous decomposition of line profiles. Unfortunately, no density sensitive line pairs formed at lower temperatures were included in these two observations. 

For the EUV jet, the total mass of the ejected material can be calculated by multiplying the mass flux density, cross section area and life time of the jet. Taking a density of log ({\it N}$_{e}$/cm$^{-3}$)=9.89, a speed of 223~km~s$^{-1}$, a jet width of $8^{\prime\prime}$ and a life time of 11 minutes \citep{YangL2011}, the total ejected mass is estimated as 5.0$\times10^{13}$~g. This is about two orders of magnitude lower than the typical value of CME mass. 

The mass of the erupted loop in the 2011 Feb 14 observation can be calculated by taking a density value of log ({\it N}$_{e}$/cm$^{-3}$)=10.01. After estimating the loop cross section area and loop length from the AIA 193\AA{} image at 19:29, the mass of the erupted loop was estimated to be 2.5$\times10^{14}$~g. Such a value is comparable to the lower limit of the typical CME mass. However, we have to bear in mind that the density value we used only represents the density of the emitting materials with a temperature around log ({\it T}/K)=6.25. 

We have to mention that both the DEM and density calculations are based on ionization equilibrium. In the case of flows, this equilibrium might be destroyed \citep{Peter2006}. To quantify how the flows impact the results of temperature and density diagnostics, further numerical simulations are needed. 

\section{Conclusion}

We have analyzed several data sets obtained by EIS during solar eruptions such as CMEs, coronal dimmings and EUV jets. We have mainly identified three types of flows and investigated the properties of them. We have also performed density diagnostics and DEM analyses for coronal dimmings, erupted CME loops and EUV jets. Our analyses suggest that spectroscopic observations can provide valuable information on the LOS kinematics and plasma properties of CMEs and EUV jets.  

Previous analyses based on single Gaussian fits reveal significant blueshift and enhanced line width in the CME-induced dimming regions. However, our detailed RB asymmetry analyses and RB-guided double Gaussian fits of the coronal line profiles clearly show blueward asymmetries in dimming regions, suggesting perhaps the presence of a relatively weak ($\sim$10\% of the total emission) high-speed ($\sim$100~km~s$^{-1}$) upflow component superimposed on a strong background emission component. This upflow component may result from the impulsive heating in the lower solar atmosphere. We have found that both the blue shift and line width correlate very well with the blueward asymmetry, suggesting that the significant blueshift and enhanced line width are actually largely caused by the superposition of the two components. This finding suggests that a small portion of the plasma in the dimming region is flowing outward at a velocity of the order of 100~km~s$^{-1}$. Our plasma diagnostics of the dimming region suggest that dimming is mainly an effect of density decrease rather than temperature change. The mass losses in dimming regions have been estimated from two different methods and they are 20\%-60\% of the masses of the associated CMEs, suggesting that a significant part of the CME mass indeed comes from the dimming region. The mass flux carried by the outflows has also been estimated from observations.

Several temperature-dependent outflows have been found immediately outside the (deepest) dimming regions. The speed increases with temperature and it reaches $\sim$150~km~s$^{-1}$ at log ({\it T}/K)=6.3. Interestingly, our RB asymmetry analysis is able to detect some of these temperature-dependent outflows. These outflows are interpreted as evaporation flows which are perhaps driven by enhanced thermal conduction or nonthermal electron beams along reconnecting field lines, or induced by the interaction between the opened field lines in the dimming region and the closed loops in the surrounding plage region. 

Profiles of emission lines formed at coronal and transition region temperatures clearly exhibit two well-separated components in erupted CME loops and EUV jets. Besides an almost stationary component accounting for the background emission, there is a highly blueshifted ($\sim$200~km~s$^{-1}$) component representing emission from the erupting material. The two components can be easily decomposed through a double Gaussian fit and we have diagnosed the electron density, performed a DEM analysis, and estimated the mass of the ejecta. Different properties of the two components suggest the importance of separating emission from different sources when studying dynamic events. Combining the speed of the blueshifted component and the projected speed of the ejecta from simultaneous imaging observations, we have calculated the real speeds of the erupted CME loop and EUV jet.

\begin{acknowledgements}
{\it SDO} is the first mission of NASA$^{\prime}$s Living With a Star (LWS) Program. EIS is an instrument onboard {\it Hinode}, a Japanese
mission developed and launched by ISAS/JAXA, with NAOJ as domestic partner and NASA and STFC (UK) as international partners. It is operated by
these agencies in cooperation with ESA and NSC (Norway). S. W. McIntosh is supported by NASA (NNX08AL22G, NNX08BA99G) and NSF (ATM-0541567, ATM-0925177). L.-D. Xia and J.-S. He are supported by the National Natural Science Foundation of China (NSFC) under contracts 40974105 and 41174148, respectively. H. Tian is supported by the ASP Postdoctoral Fellowship Program of NCAR. The National Center for Atmospheric Research is sponsored by the National Science Foundation. 
\end{acknowledgements}

\begin{appendix}
\section{Profile asymmetries not caused by blends or noise}
The high-speed outflowing component discussed in Section 3.1 is usually much weaker than the primary component, which may prompt people to think whether these high-speed outflows are caused by a weak blend or simply random noise. However, from the red contours in Figure~\ref{fig.1} we can clearly see that most Fe~{\sc{xiii}}~202.04\AA{} profiles showing significant blueward asymmetries actually have a high enough signal to noise ratio. Moreover, the blueward asymmetries form patches and they coincide with patches of significant blue shift and enhanced line width. If the blueward asymmetries are cause by random noise, we should see a random distribution of asymmetric line profiles in space. 

The cause of the blueward asymmetries by possible blends at the blue wings of the line profiles can also be ruled out, since we see these blueward asymmetries in not only one line, but all strong coronal lines in the EIS spectrum. As discussed in Section 3.1, the significant blueward asymmetries in the velocity range of 70-130~km~s$^{-1}$ at wings of these line profiles can not be caused by identified blends. 

\begin{figure*}
\centering {\includegraphics[width=0.98\textwidth]{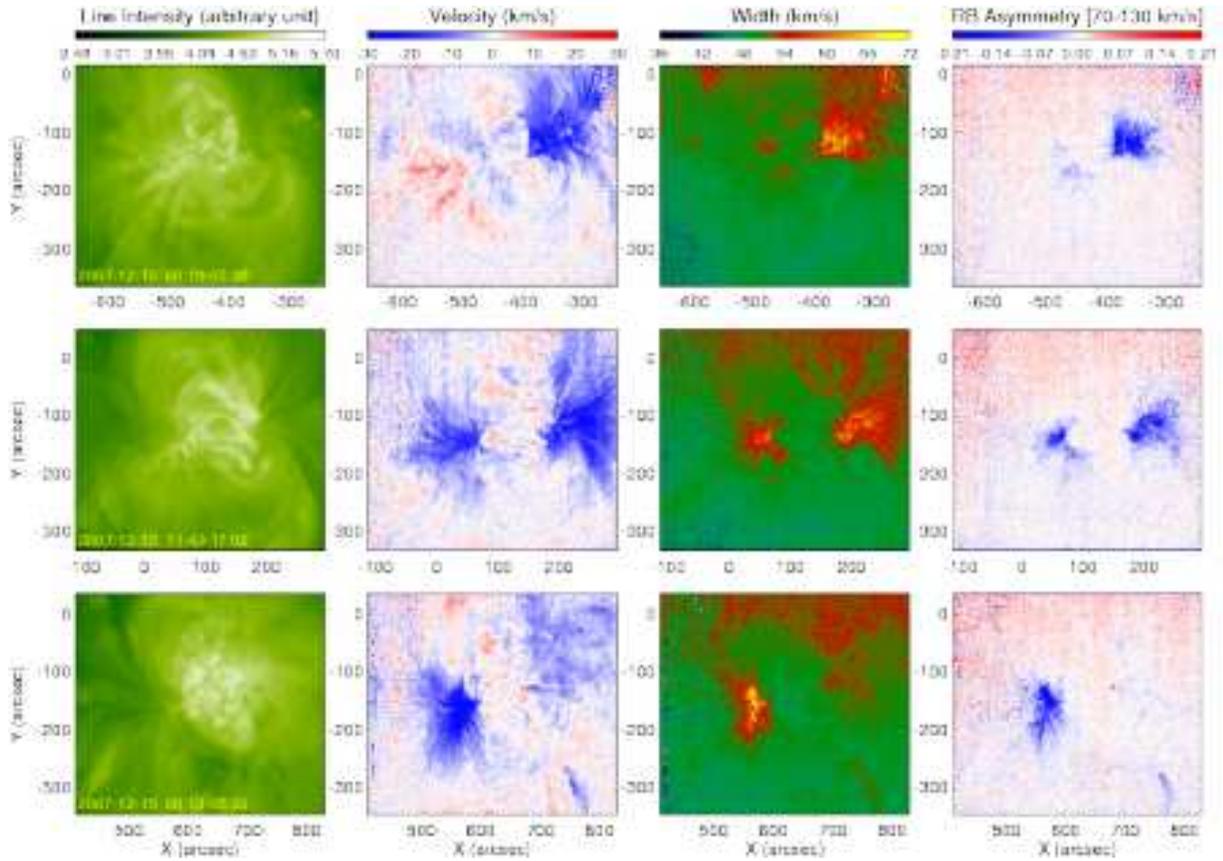}} \caption{Spatial distributions of the single Gaussian parameters and profile asymmetries (averaged over the velocity interval of 70-130~km~s$^{-1}$, as obtained from the RB$_{P}$ profiles) for Fe~{\sc{xiii}}~202.04\AA{} in a scan from 00:19 to 05:38 on 2007 Dec 10 (first row), 11:43 to 17:02 on 2007 Dec 12 (second row), and 00:13 to 05:32 on 2007 Dec 15 (third row). } \label{fig.s1}
\end{figure*}

There is another way to verify the above argument: the center to limb variation. We could not find any EIS observation of a coronal dimming region as it rotates from the limb to disk center. However, we do have EIS observations of ARs as they rotate from the limb to disk center. As discussed in Section 3.1, the high-speed outflows in dimming regions seem to be very similar to those in AR boundaries. We take three EIS raster scans of the AR 10978 from 2007 Dec 10 to 15 for an analysis. Details of these observations can be found in \cite{Bryans2010}. The AR was close to the east limb, disk center and west limb on Dec 10, 12 and 15, respectively. Figure~\ref{fig.s1} shows the spatial distributions of the single Gaussian parameters and profile asymmetries (averaged over the velocity interval of 70-130~km~s$^{-1}$, as obtained from the RB$_{P}$ profiles) for the three scans. Center to limb variations of the profile asymmetries are clearly present. When the AR was on the disk center, we see prominent blueward asymmetries at both boundaries. As the AR rotated to the west limb, the profile asymmetries disappeared at the western boundary. And the profile asymmetries almost disappeared at the eastern boundary when the AR was close to the east limb. Clearly, it is hard to explain this phenomenon by noise or blends. While our scenario, namely a high-speed outflow superimposed on a nearly static coronal background, can easily explain this center to limb variation by taking into account the LOS projection effect. When the AR was close to the west limb, the magnetic field lines in the western boundary are almost perpendicular to the LOS so that the projection of the outflow speed on the LOS direction is very small, leading to a very small velocity offset between different components and greatly reduced blueward asymmetries in the velocity interval of 70-130~km~s$^{-1}$. This effect would also reduce the magnitude of the blue shift and line width of the total emission, as also revealed by Figure~\ref{fig.s1}.   

\end{appendix}

\end{document}